\documentclass[12pt,s5paper]{article}  
\usepackage[affil-it]{authblk}
  \usepackage{fancyhdr} 
   \usepackage[latin1]{inputenc} 
  \usepackage[T1]{fontenc} 
  \usepackage{lmodern} 
  \usepackage[english]{babel} 
  \usepackage[pdftex]{graphicx} 
  \usepackage[head=25pt]{geometry} 
  \usepackage[intlimits]{amsmath} 
  \usepackage{amsthm,amssymb}
  \usepackage[usenames,dvipsnames]{color} 
  \usepackage{listings} 
  \usepackage{bold-extra} 
  \usepackage{icomma} 
  \usepackage{cancel} 
  \usepackage{url}
  \usepackage{setspace}
  \usepackage{cite}
  \usepackage[usenames,dvipsnames]{xcolor}
  \usepackage{algorithm}
  \usepackage{algorithmic}
  \usepackage{url}
  \usepackage[final]{pdfpages} 
\usepackage{graphicx}
\usepackage{caption}
\usepackage{subcaption}
 \usepackage{amsmath}
  \usepackage{makeidx}
  \usepackage{enumerate}

  \newcommand{\figref}[1]{\textbf{Figure \ref{#1}}}

  \newcommand{\secref}[1]{\textbf{Section \ref{#1}}}
  \newcommand{\figtext}[1]{\textit{\small{#1}}}

\newcommand{\RNum}[1]{\uppercase\expandafter{\romannumeral #1\relax}}

\begin{document}

\title{\textbf{\Huge{Towards Accurate Modeling of Moving Contact Lines }} }

\author{ Hanna Holmgren  \quad \quad Gunilla Kreiss}
\affil{\normalsize Department of Information Technology, Uppsala University, Box 337, SE-751 05, Uppsala, Sweden}
\date{} 
\maketitle

\begin{abstract}
A main challenge in numerical simulations of moving contact line problems is that the adherence, or no-slip boundary condition leads to a non-integrable stress singularity at the contact line. 
In this report we perform the first steps in developing the macroscopic part of an accurate multiscale model for a moving contact line problem in two space dimensions. We assume that a micro model has been used to determine a relation between the contact angle and the contact line velocity. An intermediate region is introduced where an analytical expression for the velocity exists. This expression is used to implement boundary conditions for the moving contact line at a macroscopic scale, along a fictitious boundary located a small distance away from the physical boundary.  

Model problems where the shape of the interface is constant thought the simulation are introduced. For these problems, experiments show that the errors in the resulting contact line velocities converge with the grid size $h$ at a rate of convergence $p\approx 2$.
\end{abstract}

\pagebreak
\tableofcontents
\normalsize
\pagebreak

\section{Introduction}
\begin{figure}[h!]
  \centering
    \includegraphics[width=0.55\textwidth]{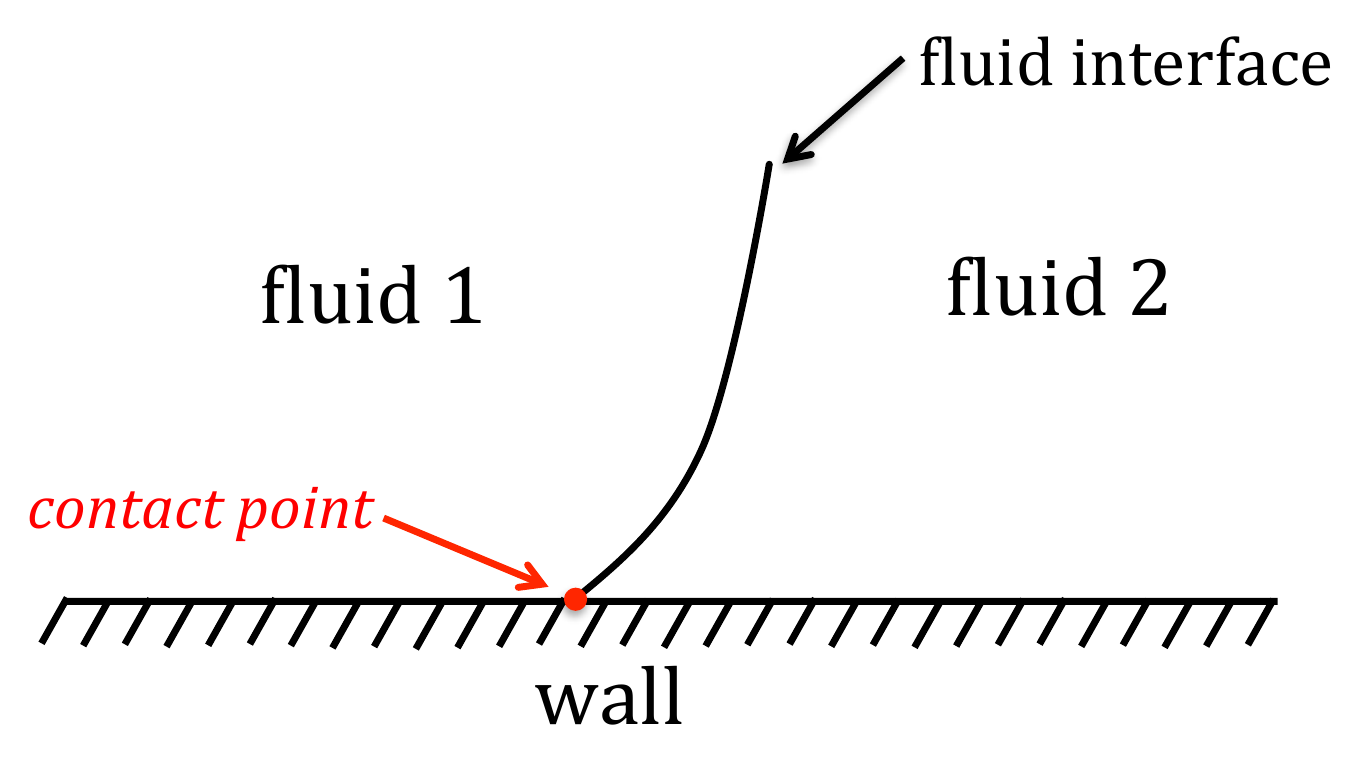}
    \caption{\figtext{Schematic illustration of a moving contact line problem in two space dimensions.}}
    \label{fig:contactline}
\end{figure}

Flow problems involving two immiscible incompressible fluids that are in contact with a solid are called moving contact line problems. The contact line is formed where the interface between the two fluids meets the solid wall. \figref{fig:contactline} depicts a schematic illustration of a contact line problem in two space dimensions, where the contact line is reduced to a contact point. In capillary and wetting flows it is the dynamics of the moving contact line that drive the flow. In such cases the physical effects that occur at the contact line are important for the overall behavior of the system \cite{REW}.
Examples of these phenomena are when a droplet is spreading on a solid surface or a liquid is rising in a narrow tube. Moving contact line problems form an important class of two-phase flows and appear both in nature and in many industrial applications \cite{REW}. Industrial applications where the contact line behavior is important include coating processes, lubrication, inkjet printing, biological flows and micro fluidics such as micropumps and so called lab-on-a-chip devices \cite{ZAHEDI, ROCCA, 1ROCCA, 2ROCCA, 3ROCCA, 4ROCCA, MARTIN}. A lab-on-a-chip devices can be used to directly diagnose a patient for different diseases by analyzing one single drop of blood, instead of sending a blood sample to a laboratory \cite{LAB-ON-A-CHIP}.

The moving contact line problem has been a subject of debate for many years. The physics governing the dynamics of moving contact lines is still not completely understood \cite{1ROCCA, HUH, REW}. The conventional hydrodynamic theory combined with a no-slip boundary condition leads to a non-integrable stress singularity at the contact line \cite{HUH, DUSSAN1}. In fact, molecular dynamics simulations show that there is some sort of fluid-wall slip in the region close to the contact line \cite{MOLEC, MOLEC2}. When performing numerical simulations of a moving contact line problem, the macroscale flow is usually the main interest. In these simulations it is too computationally demanding to resolve the problem to the microscopic scales. The singularity in the model can instead be avoided by allowing the contact line to slip \cite{MARTIN21}. A common approach is to use a so-called Navier condition with a related slip length parameter \cite{SPELT_MULTIPLE, ZARA5}. When the dynamics of the moving contact line is driving the flow only introducing slip is not enough. The effects of the microscopic dynamics of the contact line on the macroscopic flow need to be accounted for. These effects can be modeled by introducing additional conditions in the vicinity of the contact line \cite{REW}. Such conditions could for example relate the contact line velocity $U$ to the contact angle $\phi$ \cite{REW}
\begin{equation}
 U = f(\phi). 
\label{eq:CLM}
\end{equation}
These relations are usually derived with the help of microscale simulations, such as molecular dynamics simulations \cite{REN1, REN2} or simulations where the phenomenologically based phase-field model is used \cite{MARTIN2}.

Implementing slip conditions and relations on the form \eqref{eq:CLM} in an accurate way is not straight-forward. To impose slip conditions in the vicinity of the contact line different regularization techniques exist and ad-hoc procedures are common. One approach is to set the velocity in the proximity of the contact line to the given velocity from relation \eqref{eq:CLM} and then let it smoothly approach the wall velocity as the distance to the contact line increases \cite{DUSSAN, ZARA5}. Another approach is to use the Navier slip condition with a slip length parameter that is high at the contact line but reduced to zero over the next three grid points \cite{MARTIN2}. Further, to implement a contact model it is common to manipulate the representation of the interface, or the curvature of the interface, in order to follow a given contact line velocity \cite{ZAHEDI, RENARDY, MARTIN2LIU, MARTIN_BUSSMANN}. These procedures include using Dirichlet boundary conditions on equations describing the interface or fitting the interface so that is takes a given curvature \cite{MARTIN2}. 
Due to the ad-hoc regularization techniques to implement velocity boundary conditions, and/or due to the manipulation of the interface, common problems with contact line models are inaccuracies and grid effects \cite{SPELT_MULTIPLE, LUDVIG}.

In this paper we perform the first steps in constructing a higher order accurate contact line model in two space dimensions based on multiscale modeling. Here, we focus on the macroscopic part of the model. It is assume that a micro model has been used to determine a relation between the contact angle and the contact point velocity as in \eqref{eq:CLM}. We focus on accurately imposing this given contact point velocity (from the micro model) at the macroscopic scale. The choice of micro model is general (as long as it relates the contact point velocity to the contact angle).

Our macroscopic model is based on introducing an intermediate scale to the problem, in a region in the vicinity of the contact point. In this region the continuum mechanics is assumed to be valid, however the following two assumptions are made:
\begin{enumerate}
	\item The interface is essentially flat. (This is not the case if zooming in to the microscopic level where viscous bending effects are active \cite{EGGERS}. If zooming in even further molecular phenomena also have an effect on the interface shape.)
	\item The viscous effects dominate the convective and therefore the creeping flow approximation of the Navier--Stokes equations is valid.
\end{enumerate}
With these assumptions the hydrodynamic model presented in \cite{HUH} is valid. The model presented in \cite{HUH} consists of an analytical expression for the velocity close to a contact point.  This expression is here used to implement boundary conditions for the moving contact point along a fictitious boundary located a small distance away from the physical boundary. 

To model the two-phase flow dynamics we use the Navier--Stokes equations coupled to the Level set method for keeping track of the fluid interface. As a part of the level set method a reinitialization step is usually required, see \secref{sec:LS}. However, we let the implementation of an accurate reinitialization be part of the future work. Here, in the first stage of developing an accurate model we restrict ourself to study the movement of an interface that has a constant shape, and where no reinitialization is needed.

\secref{sec:theory} presents the equations used to model the two-phase flow dynamics. \secref{sec:CL_BC} describes the first steps in constructing the accurate macroscopic contact point model. Further, in \secref{sec:disc} the discretization and implementation of the two-phase models are outlined and numerical experiments are presented in \secref{sec:exp}. Finally, a summary and conclusions, including future work, are given in \secref{sec:sum}.

\section{Two Phase Flow Model}
\label{sec:theory}

\subsection{Navier--Stokes Equations}
The motion of the two immiscible fluids is given by the incompressible Navier--Stokes equations for velocity $\textbf{u}$ and pressure $p$ in non-dimensional form,
\begin{equation}
  \begin{aligned}
  \rho\frac{\partial{\bold u}}{\partial t} + \rho {\bold u} \cdot \nabla {\bold u} &= -
  \nabla p +\frac{1}{\mathrm{Re}}\nabla \cdot (2\mu\nabla^{\mathrm{s}}{\bold u})
  +\mathrm{\textbf{F}}_{st},
  \\
  \nabla \cdot {\bold u}&=0.
\label{eq:nav-stok}
  \end{aligned}
\end{equation}

Here, $\mathrm{\textbf{F}}_{st}$ is the surface tension force at the fluid interface and $\mathrm{Re}$ denotes the Reynolds number, which controls the magnitude of viscous stresses. Further, $\nabla^{\mathrm{s}}\bold{u} = \frac{1}{2}(\nabla \bold u + \nabla \bold u^T)$ denotes the rate of deformation tensor and the parameters $\rho$ and $\mu$ denote the density and viscosity measured relative to the parameters of fluid 1,
\[
\rho = \left\{\begin{array}{ll} 1 & \text{in fluid 1,} \\ \frac{\rho_2}{\rho_1} & \text{in fluid 2,} \end{array} \right. \qquad \mu = \left\{\begin{array}{ll} 1 & \text{in fluid 1,} \\ \frac{\mu_2}{\mu_1} & \text{in fluid 2.} \end{array} \right.
\]

\subsection{The Level Set Method}
\label{sec:LS}
The standard level set method \cite{LEVELSET} is used to keep track of the fluid interface and the moving contact line. The level set function $\phi(\textbf{x}, t)$ is a signed distance function and the fluid interface $\Gamma$ is given by the zero level set of $\phi$. The subdomain $\Omega_1$ occupied by fluid 1 is given by $\phi > 0$ and the subdomain $\Omega_2$ occupied by fluid 2 is given by $\phi < 0$.

The level set function is advected in time by the fluid velocity according to the following Hamilton--Jacobi equation
\begin{equation}
\frac{\partial \phi}{\partial t}+  {\bold u} \cdot \nabla \phi= 0.
\end{equation}

After advecting the fluid interface, the surface tension force $\mathrm{\textbf{F}}_{st}$ is calculated,
\begin{equation}
 \mathrm{\textbf{F}}_{st}= \frac{1}{\mathrm{We}}\kappa  {\bold n}\delta_{{ \Gamma}}, 
\end{equation}

where We is the Weber number and $\delta_{{ \Gamma}}$ is a Dirac delta function with support on $\Gamma$. The normal and curvature of the interface can be computed using the level set function,
\begin{align}
&{\bold n}=\frac{\nabla \phi}{|\nabla \phi|} \notag \\
& \kappa=-\nabla \cdot {\bold n}. \notag
\label{eq:normcurv}
\end{align}
Over time the level set function will loose its signed distance property due to discretization errors and non-uniform velocity fields. To smooth the level set function and prevent the formation of large gradients, $\phi$ has to be reinitialized with a regular interval. The standard way of doing this is to solve the following equations to steady state
\begin{equation}
\frac{\partial \phi}{\partial \tau}=-\mathrm{sign}(\phi_0)(|\nabla \phi|-1),
\label{eq:reinit}
\end{equation}
where $\phi_0$ is the level-set function before reinitialization and $\tau$ is a pseudo time step. In simulations, a regularized $\mathrm{sign}$-function that smoothly changes sign from $+1$ in the first fluid to $-1$ in the second is often used.

Performing the reinitialization in (5) does not guarantee preservation of the interface position, and thus the contact point position. Further, unphysical volume changes in the fluid phases may occur. To prevent volume changes the conservative level set method was developed in \cite{CONSLS}. However, when deriving this method contact lines were not taken into account. When applying the conservative level set method to a problem where the interface intersects the wall, we found that the computation of the curvature of the fluid interface was not accurate close to the contact line. Therefore, as a first simple study case in developing an accurate contact line model, a steady interface shape is assumed throughout the whole simulation time. In each time step the contact line position is evaluated, and the level set function is then reinitialized by re-interpolation to take the form of a signed distance function depending on the contact line position, see \secref{sec:CL_BC} for details.

\section{Contact Line Boundary Conditions}
\label{sec:CL_BC}
\begin{figure}[h!]
  \centering
    \includegraphics[width=0.8\textwidth]{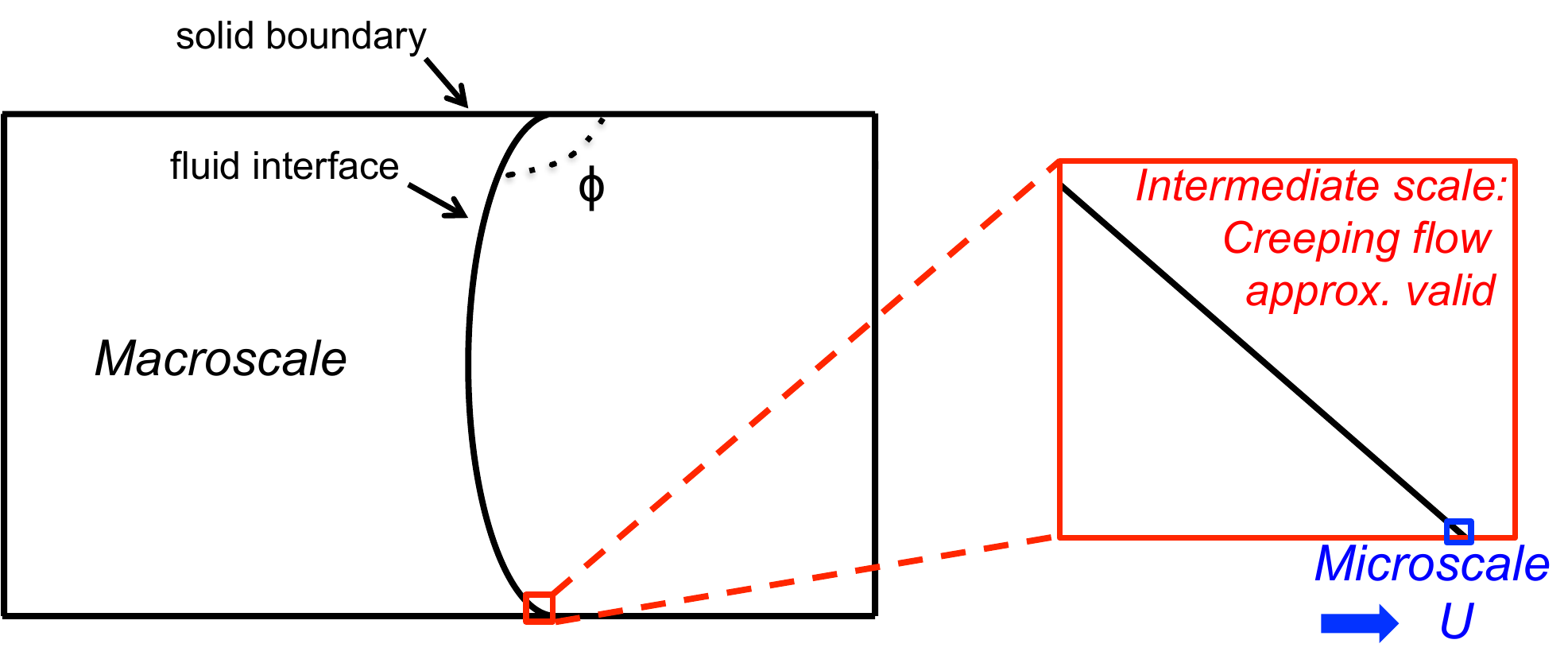}
    \caption{\figtext{Schematic illustration of the different scales in the multiscale moving contact line model.}}
    \label{fig:scales}
\end{figure}

In this report we develop the macroscopic part of a multiscale model for a moving contact line problem in two space dimensions. To model the microscopic effects of the contact line on the macroscopic level, we assume there is a relation that gives the slip velocity at the contact line $U$ as a function of the wall contact angle $\phi$ (relation \eqref{eq:CLM}). The wall contact angle $\phi$ is the angle between the interface and the wall in the macro model, see \figref{fig:scales}. This approach requires both a spatial and temporal scale separation between the local contact line behavior and global fluid flow \cite{MARTIN2}. When the dynamics of the moving contact line is driving the flow, the assumption that the flow on the microscopic scale is reacting much faster than the flow on the macroscale is justified. This temporal scale separation implies that the microscopic dynamics is in equilibrium for each wall contact angle at the macroscopic scale. Therefore no additional information from the macro model is required. These kind of relations between the contact line velocity and wall contact angle can be derived using micro models. Examples of such micro models are molecular dynamics simulation \cite{REN1, REN2} or phase field simulation \cite{MARTIN2}. The slip velocity at the contact line $U$ obtained from the microscopic relation is here used to determine velocity boundary conditions in the macroscopic model. To derive these boundary conditions an intermediate model in a region close to the contact line is introduced. 
In \figref{fig:scales} a schematic illustration of the different scales is given.

\subsection{Creeping Flow in the Intermediate Region}
\label{sec:intermediate}
To derive the macroscopic boundary conditions an intermediate region of length
scale $L$ is introduced; the red region in the schematic illustration in \figref{fig:scales}.
This region is still assumed to be in the continuum region. However, at this
length scale the viscous effects are assumed to dominate the convective, i.e. the
Reynolds number $\mathrm{Re} = \frac{\rho UL}{\mu} \ll 1$. Under these conditions the creeping flow approximation of Navier--Stokes equations is valid. As illustrated in \figref{fig:scales}, at this length scale the fluid interface could be assumed to have a flat shape. 
This assumption is not valid if you zoom in closer to the contact line, to a microscopic view. There viscous forces become very large and strongly bend the interface \cite{EGGERS}. Additionally, even closer to the contact line molecular phenomena such as diffusive mass transport affect the interface shape.

Under the assumptions of a creeping flow approximation and a flat fluid interface shape, there exists an analytical expression for the fluid velocity field in the intermediate region. The analytical model was derived in \cite{HUH} by rewriting the creeping flow approximation of Navier--Stokes equations in the form of a biharmonic equation for the stream function $\psi(r, \theta)$ in plane polar coordinates $r$ and $\theta$. The origin of the polar coordinate system is fixed to the contact line position. In terms of the stream function the polar velocity components are $v_r = -r^{-1}\frac{\partial \psi}{\partial \theta}$ and $v_{\theta} = \frac{\partial \psi}{\partial r}$. By imposing appropriate boundary and interface conditions an analytical expression for the stream function in the region close to the contact line can be derived. The analytical expression depends on the wall contact angle $0 < \phi < 180$, the contact line velocity $U$ and the viscosity ratio $Q$ and is given by
\begin{equation}
\psi(r,\theta)= r(a \sin\theta+b\cos\theta+c\theta\sin\theta+d\theta\cos\theta ), 
\end{equation}
where the coefficients $a, b, c$ and $d$ for the two different fluids, denoted by 1 and 2 below, are given by
 \begin{align}
 &a_1=-U-\pi c_1-d_1  \\
 &b_1=-\pi d_1 \\
 &c_1=US^2[S^2-\gamma\phi+Q(\phi^2-S^2)]/D   \\
 &d_1=USC[S^2-\gamma\phi+Q(\phi^2-S^2)-\pi \tan\phi]/D \\
 &a_2=-U-d_2  \\
 &b_2=0 \\
 &c_2=US^2[S^2-\gamma^2+Q(\delta\phi-S^2)]/D  \\
 &d_2=USC[S^2-\gamma^2+Q(\delta\phi-S^2)-Q\pi \tan \phi]/D  ,
 \end{align}
where 
\begin{align}
&S=\sin\theta \notag \\
&C=\cos \theta \notag \\
&\gamma=\phi-\pi \notag \\
&Q=\mu_A/\mu_B \notag \\
&D=(SC-\phi)(\gamma^2-S^2)+Q(\delta-SC)(\phi^2-S^2) .\notag 
\end{align}

The resulting analytical velocity field constitutes a similarity solution since it is independent of the distance to the contact line $r$. As mentioned above, the analytical model is not valid in the immediate neighborhood of the contact line, at the microscopic length scales. In fact the analytical velocity field is discontinuous at the contact line and shear stresses, pressure, and viscous dissipation rate increase without bound as the contact line is approached. Nevertheless, the model can be used do describe the dynamics in the intermediate region: ``the model may approximate reality well in a slightly removed region where the fluid interface is substantially flat and the flow qualifies as creeping'', \cite{HUH}.

\begin{figure}[h!]
  \centering
    \includegraphics[width=0.8\textwidth]{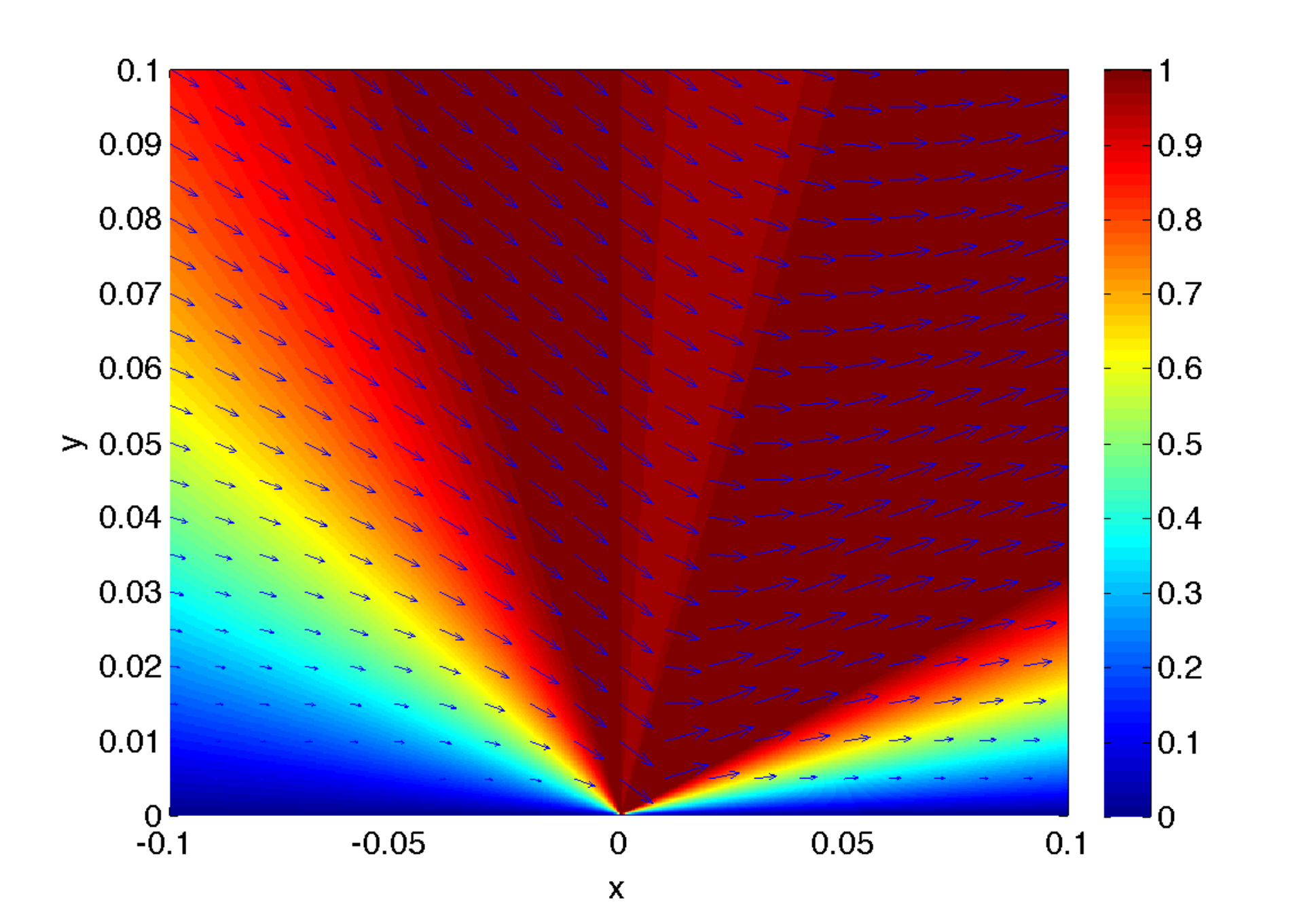}
    \caption{\figtext{The analytical velocity field in the intermediate region for $\phi = 45$, $U = 1$ and $Q=1$.}}
    \label{fig:simsol}
\end{figure}

\begin{figure}[h!]
  \centering
    \includegraphics[width=0.5\textwidth]{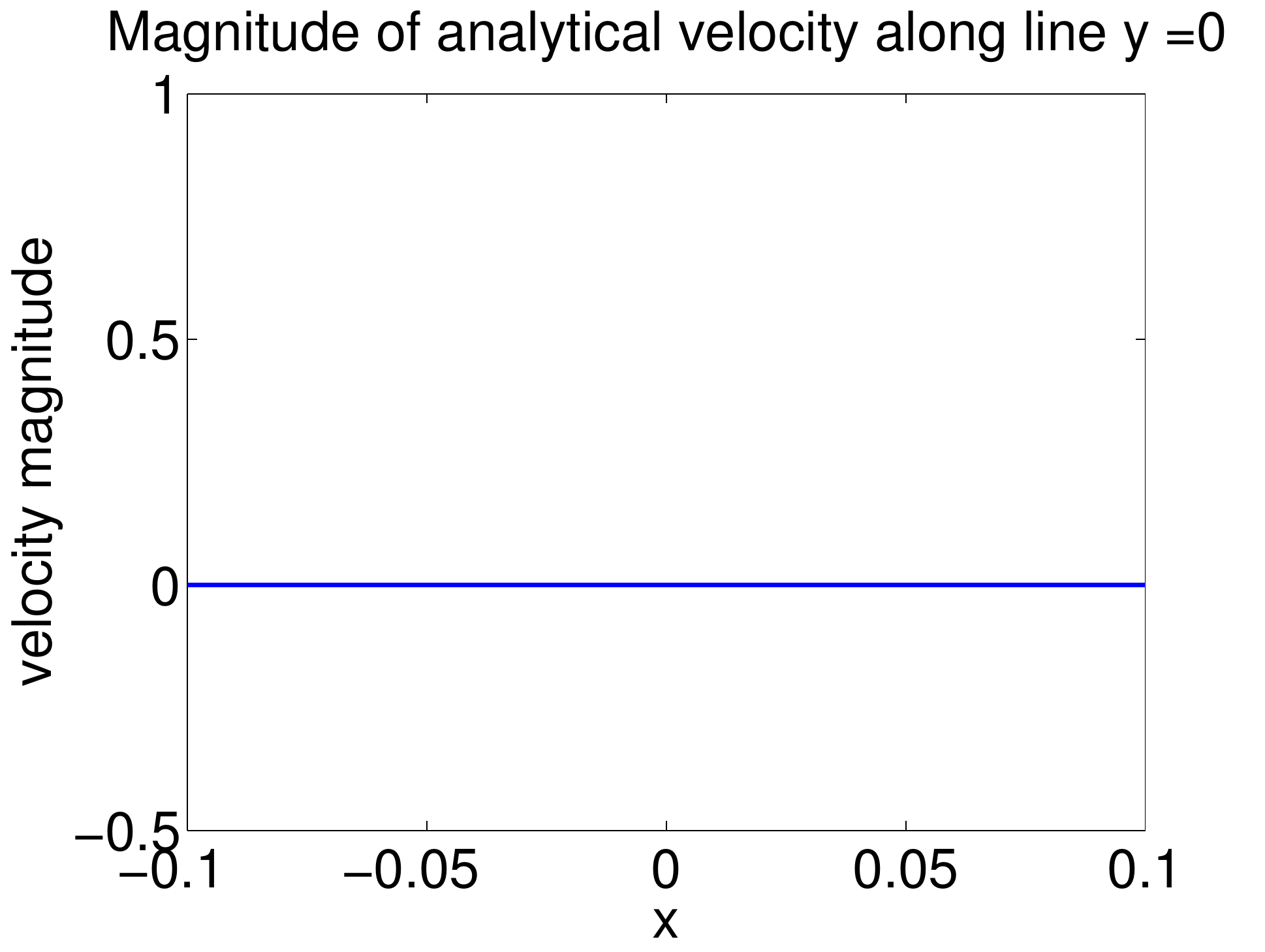}
    \includegraphics[width=0.5\textwidth]{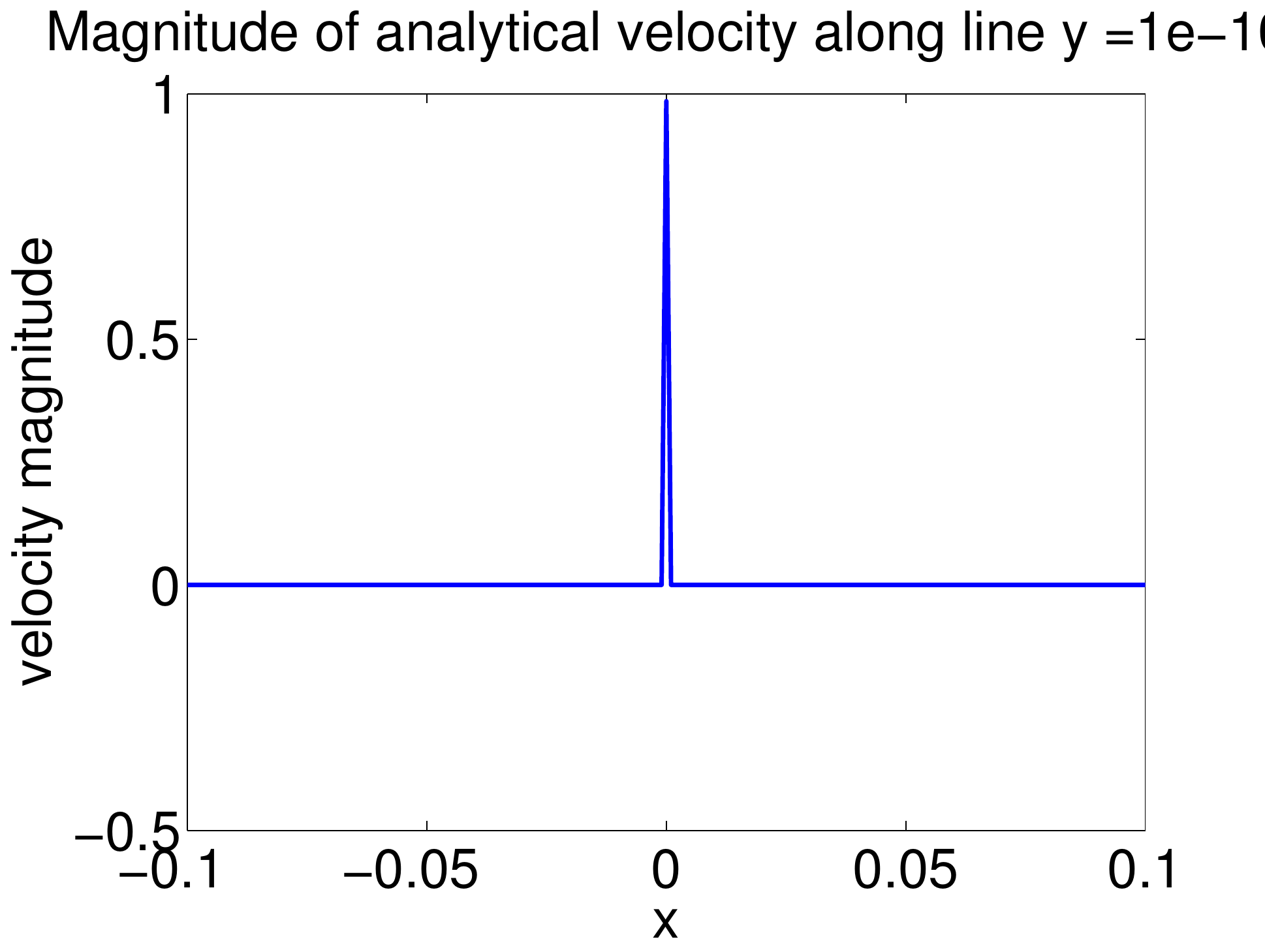}
    \includegraphics[width=0.5\textwidth]{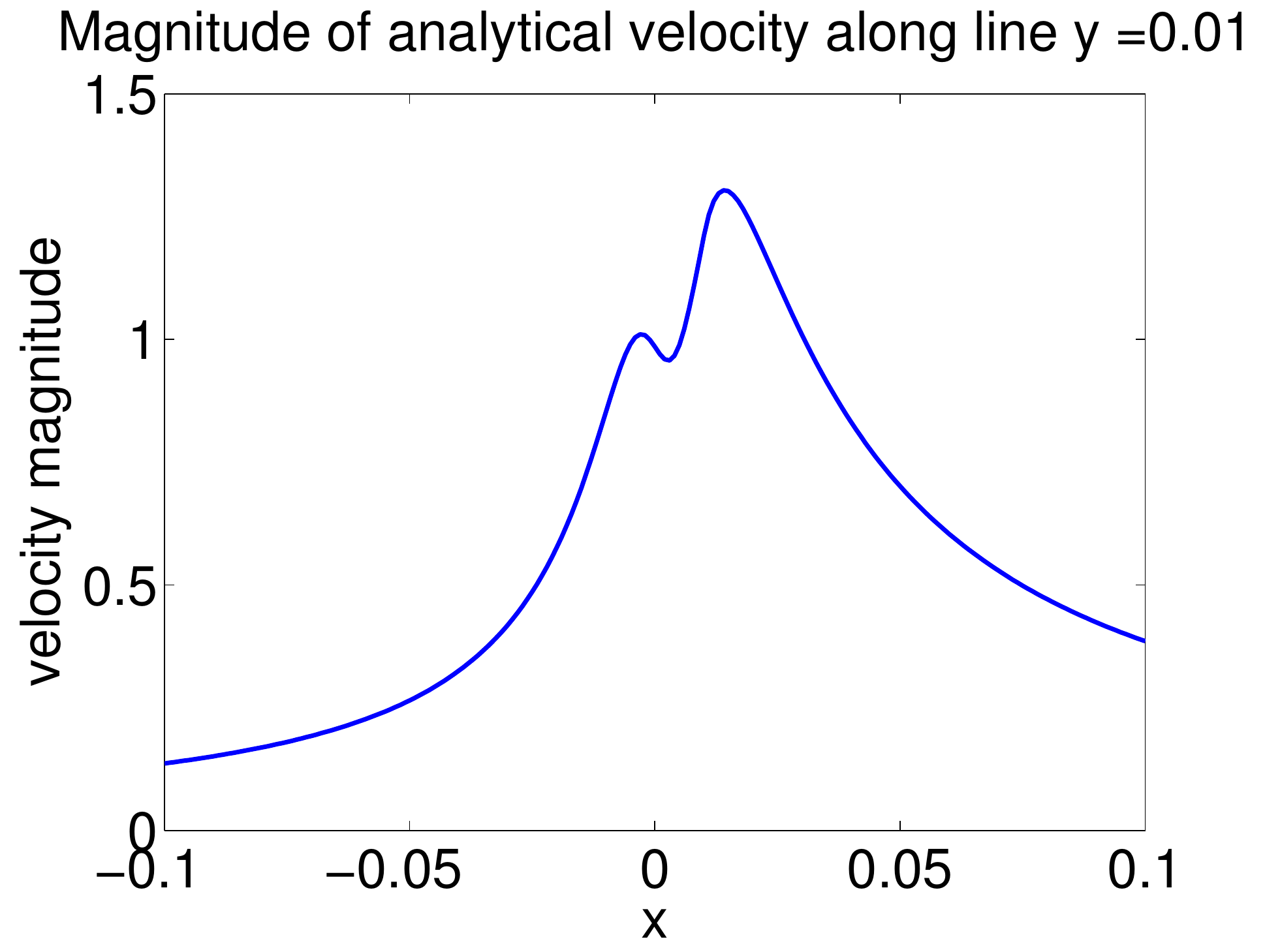}
    \caption{\figtext{The analytical velocity field in the intermediate region in \figref{fig:simsol} evaluated along three lines $y = 0$, $y = 10^{-10}$ and $y = 10^{-2}$.}}
    \label{fig:lines}
\end{figure}

In \figref{fig:simsol} the magnitude of the analytical velocity in the intermediate region (the red box in \figref{fig:scales}) is plotted for the case with a contact angle $\phi = 45$, contact velocity $U = 1$ and viscosity ratio $Q = 1$. In \figref{fig:lines} the magnitude of the velocity along the three lines $y = 0$, $y = 10^{-10}$ and $y = 10^{-2}$ from the domain in \figref{fig:simsol} are plotted. It can be seen that the velocity is zero along the whole solid boundary, i.e. along the whole line $y = 0$, also at the contact line. However, the second plot in \figref{fig:lines} depicts that just inside the boundary the velocity is non-zero in the vicinity of the contact point. This illustrates the velocity discontinuity at the contact point. The further away from the wall, the wider is the peak in the velocity at the contact point along a line $y = \delta$, see the third plot in \figref{fig:lines} for example. More precisely, since the analytical velocity field constitutes a similarity solution, the width of the peak increase linearly with increasing $\delta$.

\subsection{Velocity Boundary Conditions}
\label{sec:VEL_BC}
\begin{figure}[h!]
  \centering
    \includegraphics[width=0.7\textwidth]{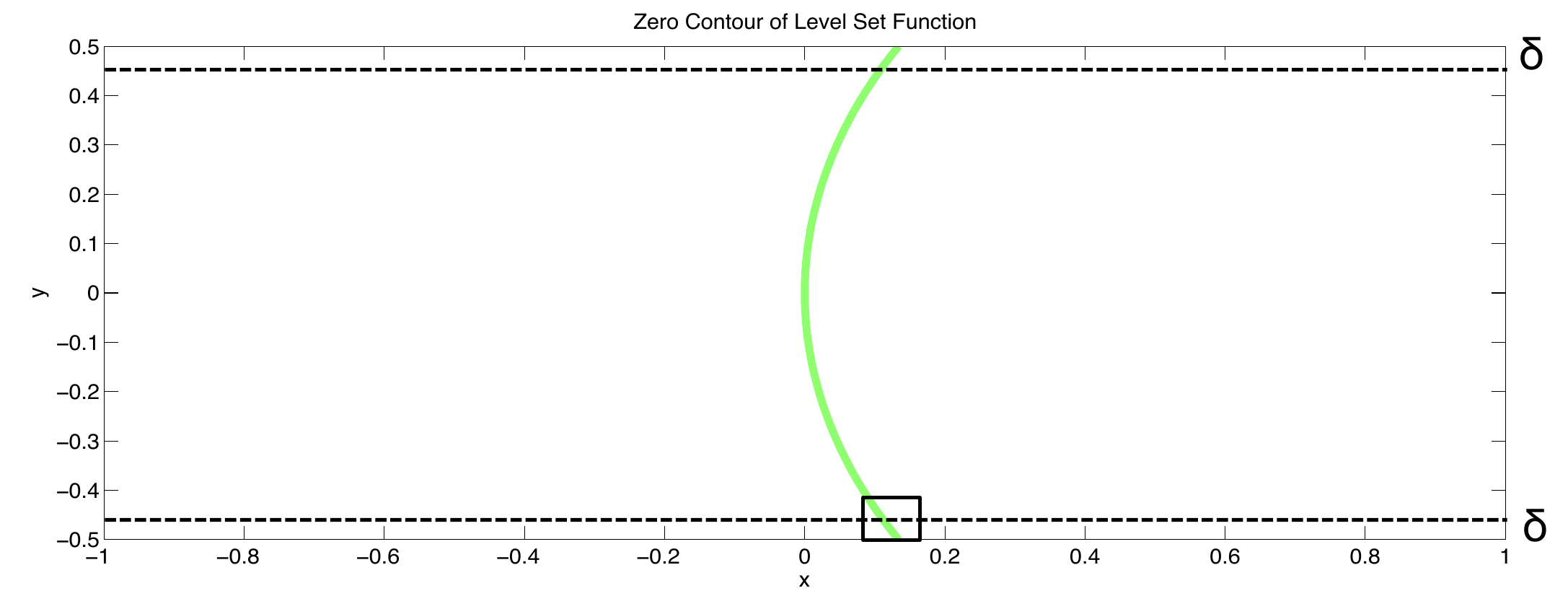}
    \caption{\figtext{Modified domain.}}
    \label{fig:mod_dom}
\end{figure}

\begin{figure}[h!]
  \centering
    \includegraphics[width=0.85\textwidth]{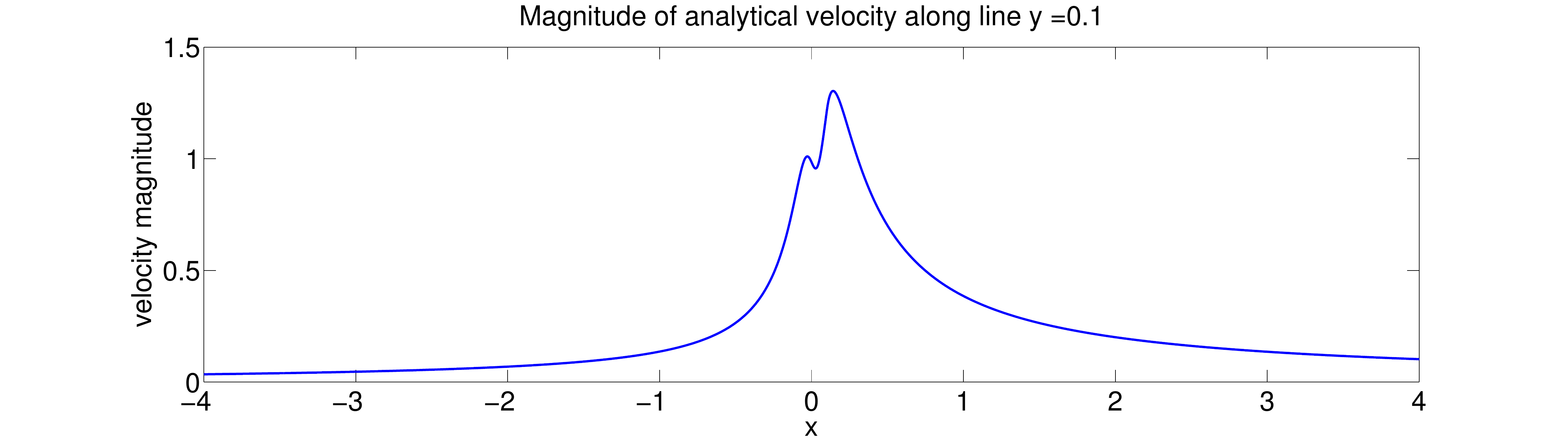}
    \caption{\figtext{Velocity Dirichlet boundary condition derived from the analytical intermediate model.}}
    \label{fig:fullline}
\end{figure}

The analytical model in the intermediate region described above, is here used to develop a velocity boundary condition for the macroscopic simulation. To avoid the singularity at the contact point in the analytical model, parts of the intermediate region will not be included in the macroscopic simulation. 
Consequently, a computational domain that is $2\delta$ smaller in the direction perpendicular to the solid wall will be used, see \figref{fig:mod_dom}. Along the new fictitious boundary, which is $\delta$ inside the physical boundary, we impose the analytical velocity from the intermediate model as a Dirichlet boundary condition for the velocity in the macroscopic model. If we use the domain in \figref{fig:mod_dom} with $\delta = 0.1$ for example, the velocity function in \figref{fig:fullline} is used along the fictitious boundary. The information about the slip contact point velocity $U$ from the microscopic model, i.e. the information about the movement of one single point, is transformed into a velocity boundary condition along the whole (fictitious) boundary.

\subsubsection{Choosing $\delta$ and Spatial Grid Size $h$}
\label{sec:delta}
\begin{figure}[h!]
  \centering
    \includegraphics[width=1\textwidth]{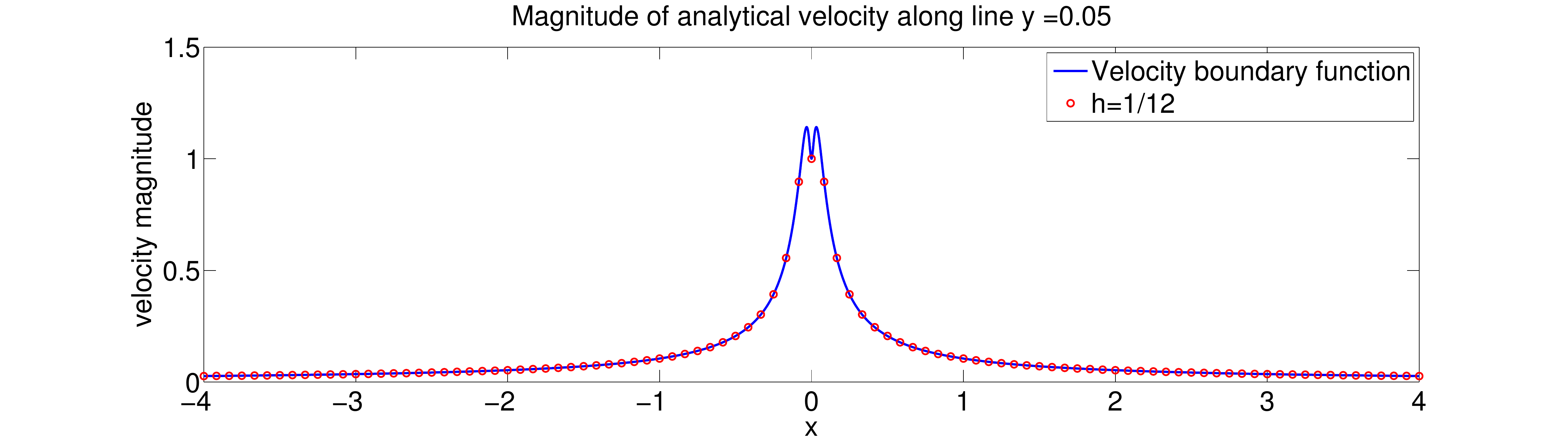}
    \includegraphics[width=1\textwidth]{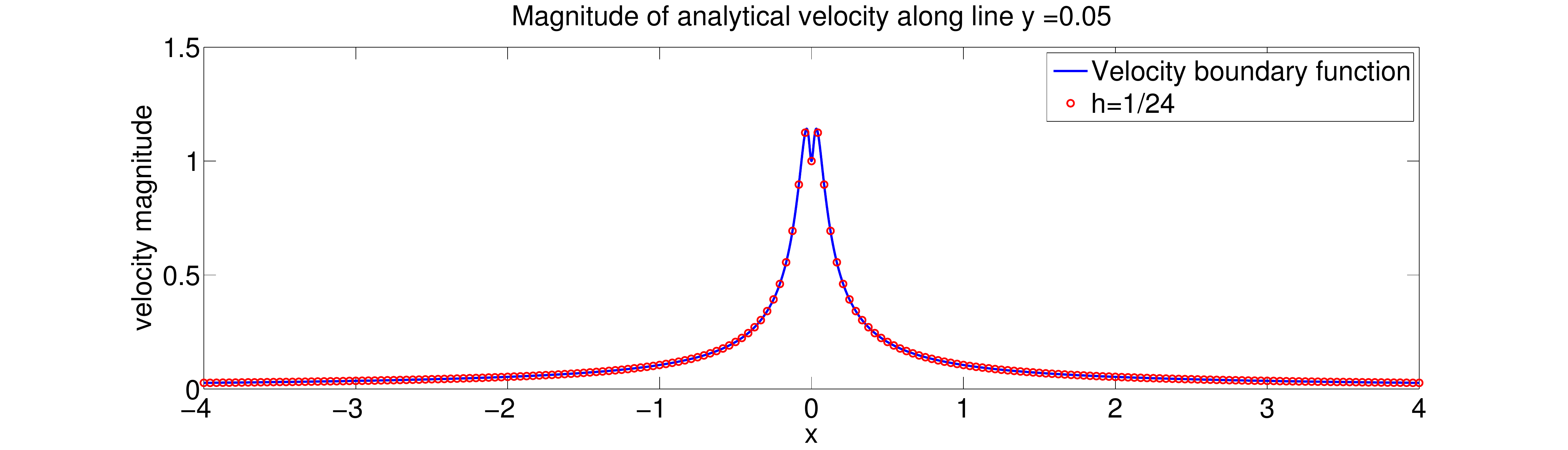}
    \caption{\figtext{Velocity boundary function and corresponding values at grid point for $h = 1/12$ and $h=1/24$ ($\phi=90$, $U =1$, $Q=1$).}}
    \label{fig:delta}
\end{figure}

When implementing the velocity boundary condition described above, care has to be taken when choosing the value of $\delta$. The creeping flow approximation is valid in a region with length scale $L$ (intermediate region) and $\delta < L$ is required. However, the smaller $\delta$ the sharper is the peak in the boundary velocity function at the contact point (\figref{fig:lines}). Therefore, when discretizing in space the grid must sufficiently resolve the features of the boundary function. To investigate what grid size $h$ is sufficient for a certain $\delta$, it is instructive to plot the resulting velocity function together with the corresponding discrete version of the boundary function. In \figref{fig:delta} for example the resulting velocity boundary function for $\delta = 0.05$ and contact angle $\phi = 90$, contact point velocity $U = 1$ and viscosity ratio $Q = 1$ is plotted together with corresponding discrete versions where $h = 1/12$ and $h = 1/24$. There it can be seen that only the smaller grid size will be able to capture details of the peak in the boundary function. Since the width of the peak in the velocity function depends linearly on $\delta$ one knows how much the grid needs to be refined when decreasing $\delta$, assuming the limit of the grid size for a certain $\delta$ is known.

\subsection{Interface Boundary Conditions}
Forcing the interface to move according to the contact point velocity using Dirichlet boundary conditions might lead to incompatibility of the contact line location between the velocity field and the interface representation via the level set method. Further, advecting the contact point explicitly using Dirichlet boundary conditions for the level set function might effect the mass conservation. Since the velocity boundary condition developed in previous subsection depends on the contact point velocity $U$ (given from the micro model), the interface close to the contact point should automatically be advected with that given velocity. Thus, there is no need for explicitly moving the interface using additional Dirichlet boundary conditions. Along the wall we simply use homogenous Neumann boundary conditions for the level set function. Possible oscillations in the level set function due to inflowing characteristics is smoothed during the reinitialization.

\section{Discretization and Implementation}
\label{sec:disc}
For the implementation we use the existing two-phase flow solver described in \cite{MARTINBABA} with suitable modifications to account for moving contact lines. The conservative level set method implemented in \cite{MARTINBABA} is also changed to the standard level set method (described in \secref{sec:LS}). Additionally, the order in which the equations are solved in each time step is changed so that the Navier--Stokes equations are solved before the interface is advected (and not the other way around as in \cite{MARTINBABA}). The solver is implemented in the C++ based Finite Element open source library deal.ii \cite{DEAL1, DEAL2}. The equations in \secref{sec:theory} are discretized in space using the Finite Element Method. For the level set function piecwise continuous linear shape functions on quadrilaterals, i.e. $Q_1$ elements, are used. For the incompressible Navier--Stokes equations we use the Taylor--Hood elements $Q_2Q_1$, i.e., shape functions of degree two for each component of the velocity and of degree one for the pressure. With these elements the Babu\^ska--Brezzi (inf--sup) condition \cite{INFSUP} is fulfilled in order to guarantee the existence of a solution. Finite element discretizations of equations of transport type, such as the level set equation, typically need to be stabilized. Here however, no stabilization is used since the reinitialization will take care of possible oscillations. Dirichlet boundary conditions are imposed strongly.

For time stepping, each of the level set equation and Navier--Stokes equations are discretized using the second order accurate, implicit BDF--2 scheme. In order to avoid an expensive coupling between the incompressible Navier--Stokes part and the level set part (via the variables $\textbf{u}$ and $\phi$) a temporal splitting scheme is introduced. In order to maintain second oder accuracy in time, at each time step $n$ an estimate of the level set function is extrapolated from the values at time steps $n-1$ and $n-2$. This estimate is used to evaluate an approximation of the surface tension force. With this surface tension force, the BDF-2 time step for the Navier--Stokes equations is then performed. Finally, the level set function is propagated in time, according to the velocity $\textbf{u}^n$ obtained from the Navier--Stokes step, again using the BDF-2 method. 
The splitting between the level set and Navier--Stokes parts corresponds to an explicit treatment of surface tension, which gives rise to a time step limit
\begin{equation}\label{eq:capillary_time_limit}
  \Delta t \leq c_1 \frac{\mathrm{We}}{\mathrm{Re}}h + \sqrt{\left(c_1\frac{\mathrm{We}}{\mathrm{Re}}h\right)^2 + c_2 \mathrm{We}\, h^3}, 
\end{equation}
where $c_1$ and $c_2$ are constants that do not depend on the mesh size $h$ or the material parameters. For more details about the time discritization see \cite{MARTINBABA}.

After time discretization and linearization of the Navier--Stokes equations using the implicit Newton method, linear systems need to be solved. For the level set equation, a BiCGStab solver \cite{SAAD} is used due to non-symmetry. The resulting system after discretization of the Navier--Stokes equations is of saddle point structure and solved by an iterative GMRES solver \cite{SAAD}. For preconditioning, a block-triangular operator constructed using the so called Schur complement of the block system is applied from the right \cite{ELMAN}. Most of the iterative solvers spend the bulk of the computing time in matrix-vector products. Therefore, the fast matrix-free methods from \cite{KORMANN, KKORMANN} based on cellwise integration are used for matrix-vector products. This enables matrix-free matrix-vector products that are up to an order of magnitude faster on $Q2$ elements. For more details about the linear solvers and matrix-free methods used, we again refer to \cite{MARTINBABA}.

\section{ Numerical Experiments}
\label{sec:exp}
To investigate the moving contact line boundary condition developed in \secref{sec:CL_BC} we use three test problems. All test problems are performed on a two dimensional channel with a length of 10 and a hight of 2, and the initial contact point position at $x = -0.5$. Further, in all simulations both fluids are assumed to have viscosity $\mu = 0.7$ and density $\rho = 1$. We start with a very simple set up in the first test problem, and proceed by adding more complexities into the two preceding test problems. At the stage of this report we do not use a specific relation between the contact angle and velocity from a specific micro model, instead we construct hypothetical examples for the relation between the contact point velocity U and wall contact angle $\phi$.

For all problems the gradient of the level set function, $\nabla \phi$, gets distorted at the interface close to the contact point if no reinitialization is used. 
If very steep or very flat gradients are formed, the evaluation of the zero contour, the normal vectors, and the curvature accuracy becomes an issue. 
However, as mentioned in \secref{sec:LS} performing the standard reinitialization in \eqref{eq:reinit} will modify the contact point position. Therefore, as a first simple study case to validate the contact line boundary conditions derived in last section, a steady interface shape is assumed throughout the whole simulation and ``reinitialization'' is performed in the end of each time step according to the following steps:
\begin{enumerate}
	\item The contact point position is evaluated by cubic interpolation using the level set function values at the two degrees of freedoms closest to the contact point in each direction along the wall.
	\item The level set function values in the whole domain are redefined to represent a distance function according to the constant interface shape (what shape depends on the problem set up) with contact point position according to the calculated position in the previous step.
\end{enumerate}

We are interested in investigating the ability of the contact line boundary conditions to accurately advect the contact point according to the velocity $U$ obtained from the micro model. Therefore, it is important that the evaluation of the contact point position in the first step presented above is accurate, why a cubic interpolation is used. 
However, the procedure to modify the interface to take a fixed form introduces an error, in the end of each time step, that does not depend on the time step size $\Delta t$. To investigate spatial grid convergence the time step is therefore fixed to $\Delta t = 0.01$ for all simulations. The spatial grid sizes used are $h = 1/24,1/32,1/40,1/48,1/56$ and with the value of $\Delta t = 0.01$, the spatial discretization error is assumed to dominate the temporal error (for all spatial grid sizes).

\subsection{Flat Fluid Interface with Creeping Flow Velocity Field}
\begin{figure}[h!]
  \centering
    \includegraphics[width=1\textwidth]{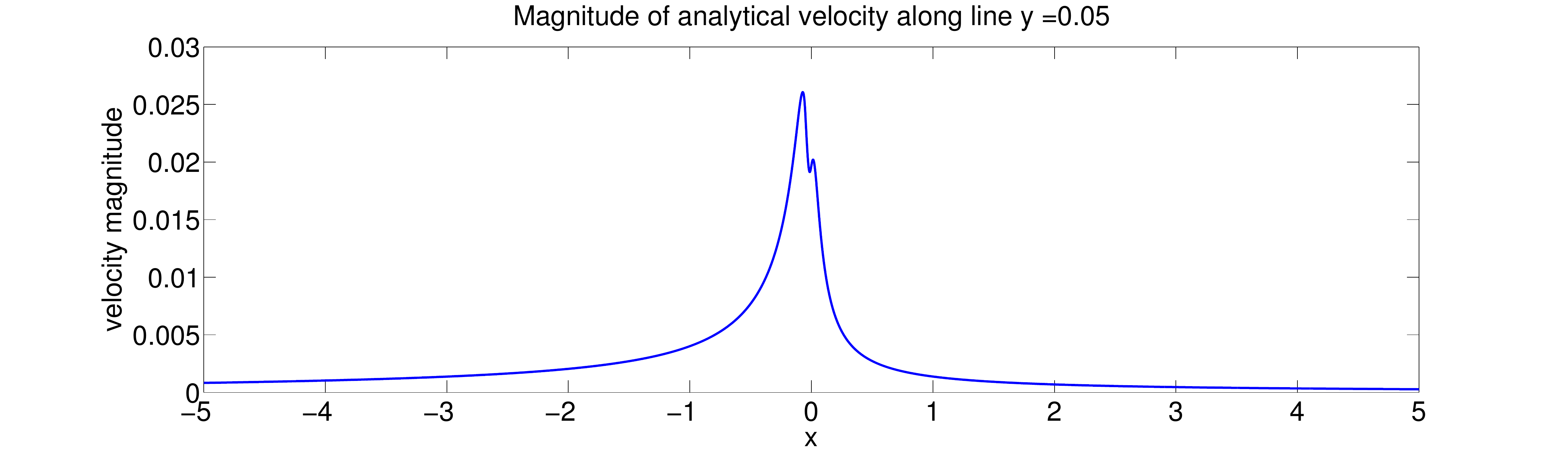}
    \caption{\figtext{Velocity boundary function at $\delta = 0.05$ for the first test problem: $\phi = 135$, $U =0.02$,  $Q=1$.}}
    \label{fig:BC_FN1}
\end{figure}

\begin{figure}[h!]
  \centering
    \includegraphics[width=1\textwidth]{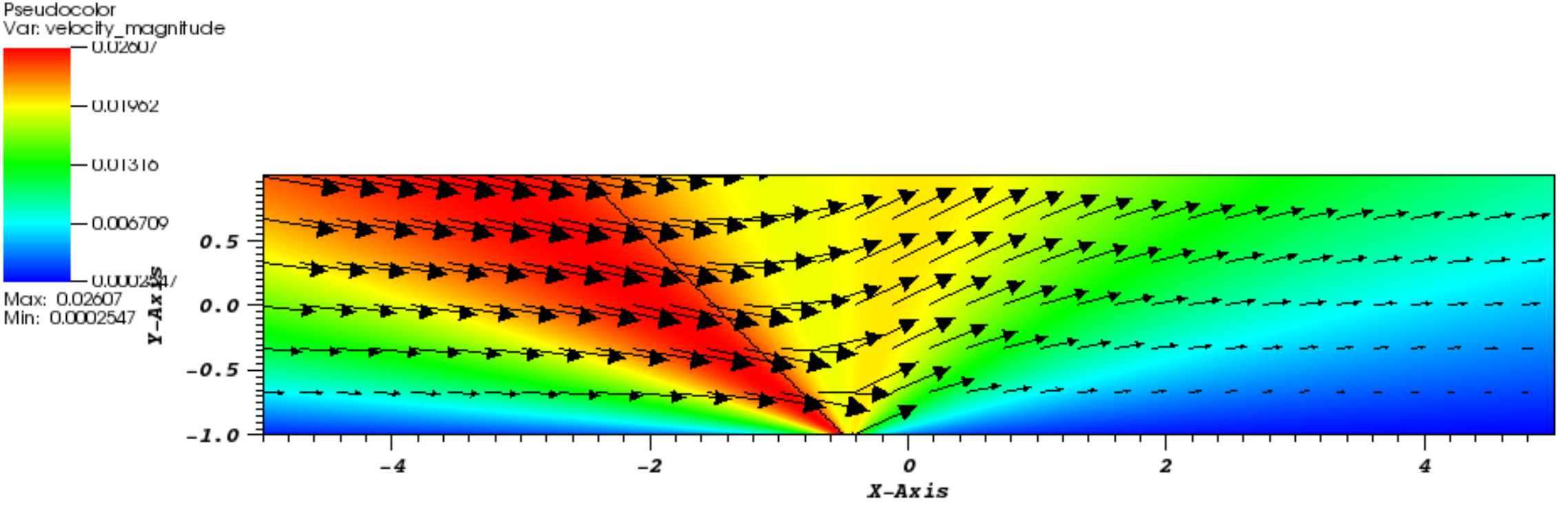}
    \caption{\figtext{ Initial configuration of the first test problem: $\phi = 135$, $U = 0.02$, $Q = 1$. The black line represents the zero level set of the level set function, i.e. the fluid interface.}}
    \label{fig:initial1}
\end{figure}

The first test problem consists of simple advection of a flat interface using a given velocity field (i.e. the Navier--Stokes part is not solved for). The given velocity field is the analytical velocity field from the creeping flow approximation in the intermediate model described in \secref{sec:intermediate}. The wall contact angle of the flat interface is taken to be $\phi = 135$ and we hypothetically relate this angle to a contact point velocity of $U = 0.02$.
As explained in \secref{sec:delta}, when choosing $\delta$ we need to make sure $\delta < L$ where
$L$ is the characteristic length scale of the intermediate region close to the contact
line where the creeping flow approximation is valid. For the creeping flow to be
valid we need $\mathrm{Re} = \frac{\rho UL}{\mu} \ll 1$ or $L \ll \frac{\mu}{\rho U}$ which for this model problems implies
the following condition on the distance to the physical boundary: $\delta < L \ll 35$ and we use $\delta = 0.05$. \figref{fig:BC_FN1} shows the resulting velocity boundary function (see \secref{sec:VEL_BC}). Further, \figref{fig:initial1} illustrates the initial configuration of the first test problem.

\begin{figure}[h!]
  \centering
    \includegraphics[width=0.6\textwidth]{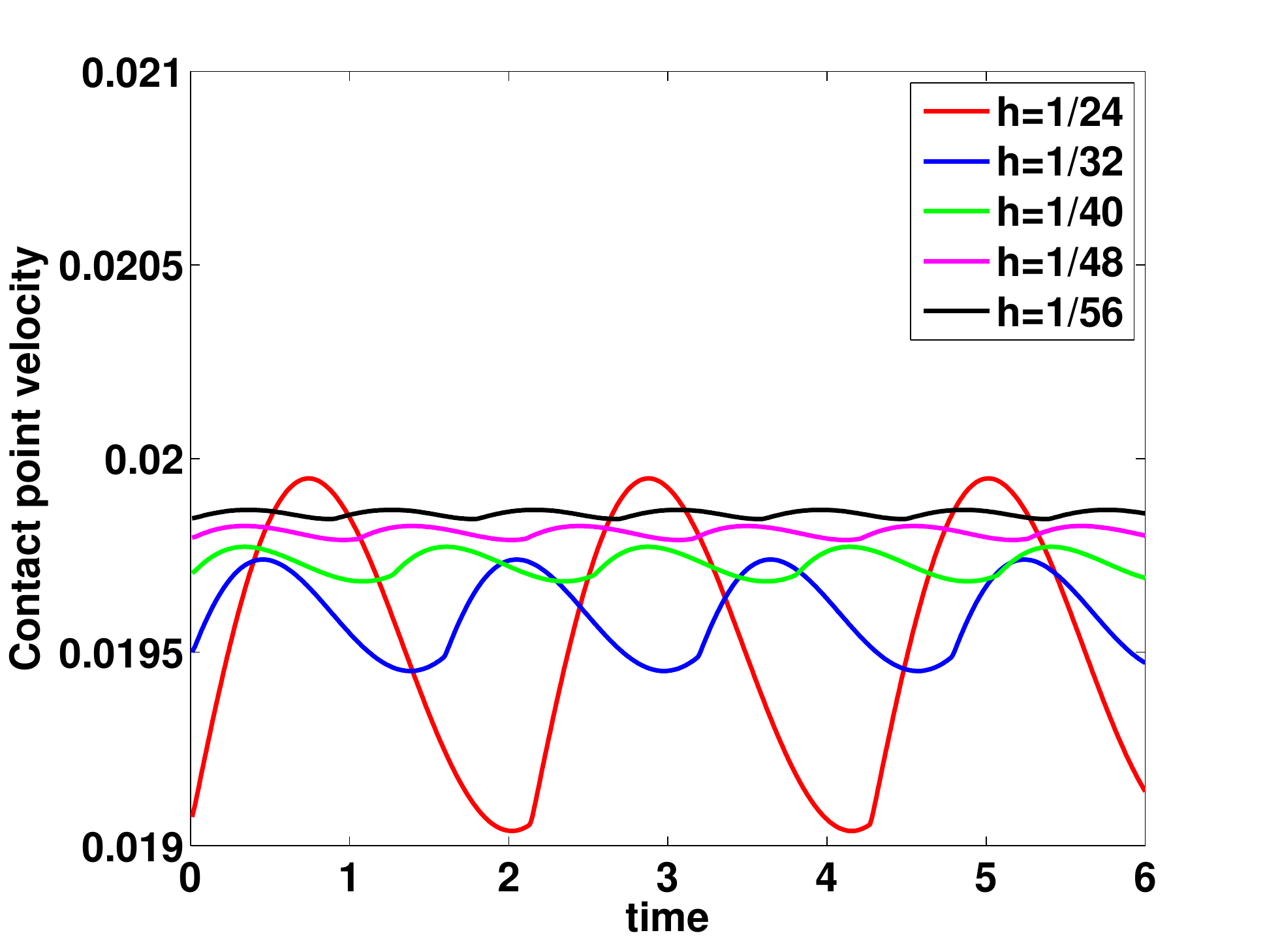}
    \caption{\figtext{Resulting contact point velocity for the first model problem.}}
    \label{fig:velocities1}
\end{figure}

\begin{figure}[h!]
  \centering
    \includegraphics[width=0.6\textwidth]{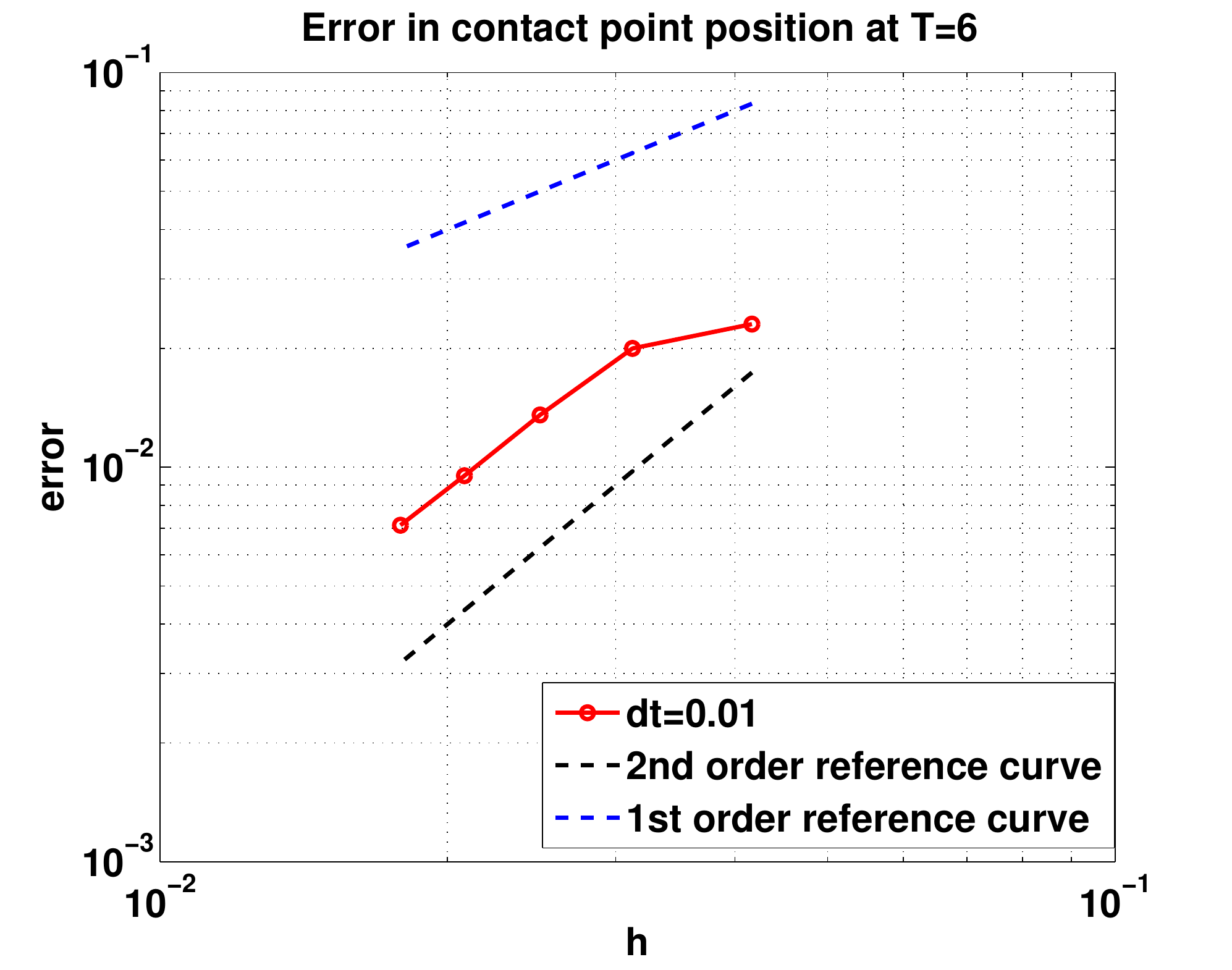}
    \caption{\figtext{ Error in contact point position at T = 6 for the first model problem.}}
    \label{fig:conv1}
\end{figure}

Using the procedure described in \secref{sec:delta}, we find an upper limit of the spatial grid size h of 1/24. Thus, we perform simulations using the different grid sizes $h = 1/24,1/32,1/40,1/48,1/56$ for a non-dimensional total time $T = 6$. The resulting contact point velocity in each time step is plotted as a function of time in \figref{fig:velocities1}. The period of the oscillations corresponds to the time it takes for the contact line to pass one grid cell.  
At time $t = T$ we measure the relative error in the contact point position compared to the correct position $UT$ for the different grid sizes. In \figref{fig:conv1} it can be seen that grid convergence is obtained with a rate of convergence $p \approx 2$. In \figref{fig:conv1} it can be seen that the grid size $h = 1/24$ is probably not capable of resolving the velocity boundary function enough, since the rate of convergence is higher for the smaller grid sizes.

\subsection{Flat Fluid Interface Coupled to Navier--Stokes Equations}
\begin{figure}[h!]
  \centering
    \includegraphics[width=1\textwidth]{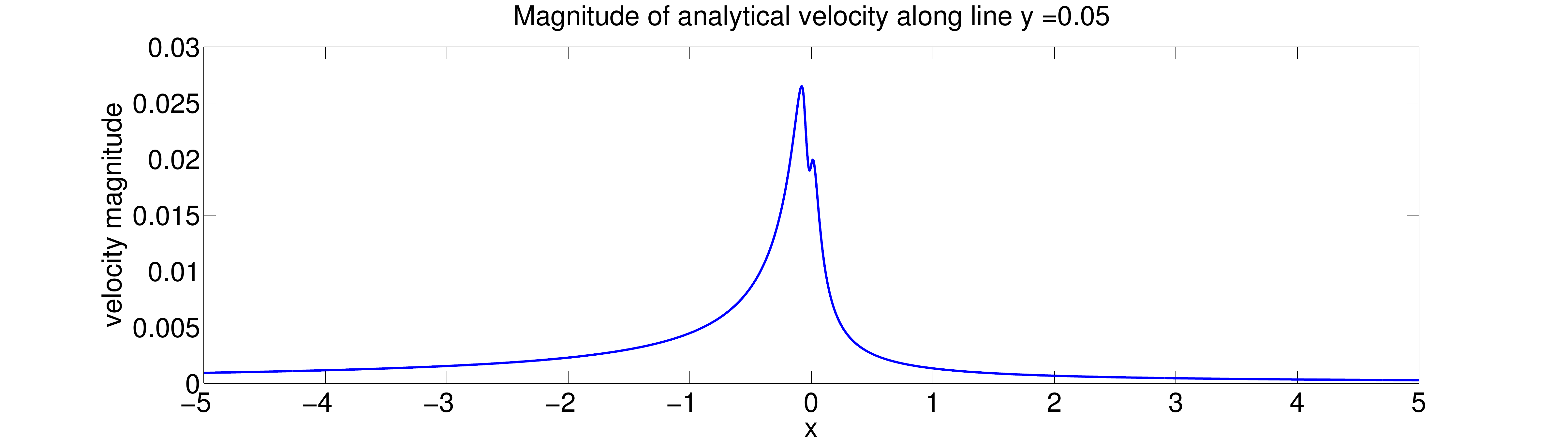}
    \caption{\figtext{Velocity boundary function at $\delta = 0.05$ for the first test problem: $\phi = 140$, $U =0.02$,  $Q=1$.}}
    \label{fig:BC_FN2}
\end{figure}

In next step, advection of an interface having a constant flat shape is still studied, but with the underlying velocity field coming from the Navier--Stokes solution, i.e. all equations in \secref{sec:theory} are solved for. The wall contact angle of the flat interface is taken to be $\phi = 140$ and we hypothetically relate this angle to a contact point velocity of $U = 0.02$. Again, we choose $\delta = 0.05$ and we find an upper limit of the spatial grid size $h$ to be 1/24. The resulting velocity boundary function for the second test problem is showed in \figref{fig:BC_FN2}.

\begin{figure}[h!]
  \centering
    \includegraphics[width=1\textwidth]{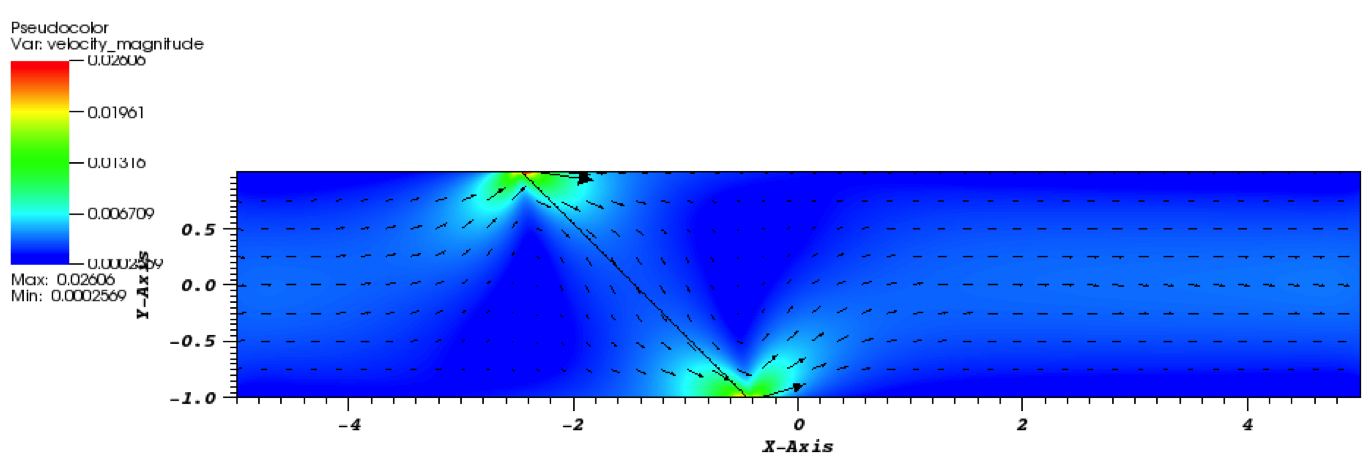}
    \caption{\figtext{Resulting velocity field for the second test problem: $\phi = 140$, $U = 0.02$, $Q = 1$.}}
    \label{fig:result2}
\end{figure}

Simulations with the different grid sizes are performed and the resulting velocity field at $T = 6$ for the grid size $h = 1/56$ is plotted in \figref{fig:result2}. Since the curvature of the interface is zero, the surface tension force in the Navier--Stokes equations is zero and the velocity field is therefore not affected by the interface. Further, the flat fluid interface is not realistic why we use a circular fluid interface shape in the third and final test problem.

\begin{figure}[h!]
  \centering
    \includegraphics[width=0.5\textwidth]{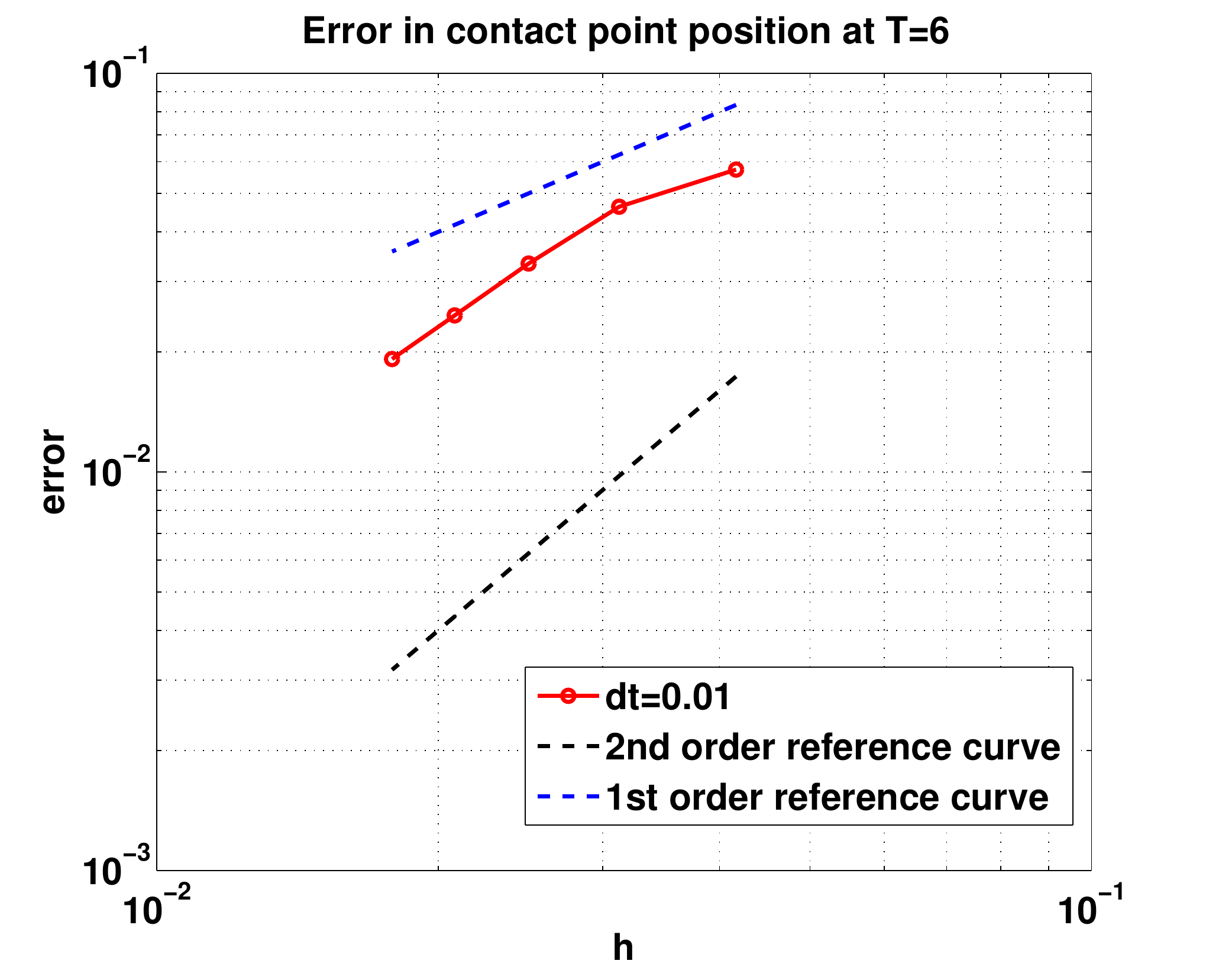}
    \caption{\figtext{ Error in contact point position at T = 6 for the second model problem.}}
    \label{fig:conv2}
\end{figure}

However, the second model problem is used for studying the the grid convergence of the contact line boundary conditions in the Navier--Stokes equations. So, at time $t = T$ we measure the relative error in the contact point position compared to the correct position $UT$ for the different grid sizes and the result is shown in \figref{fig:conv2}. Again, it can be seen that grid convergence is obtained with a rate of convergence $p \approx 2$.

\subsection{Circular Fluid Interface Coupled to Navier--Stokes Equations}
\begin{figure}[h!]
  \centering
    \includegraphics[width=1\textwidth]{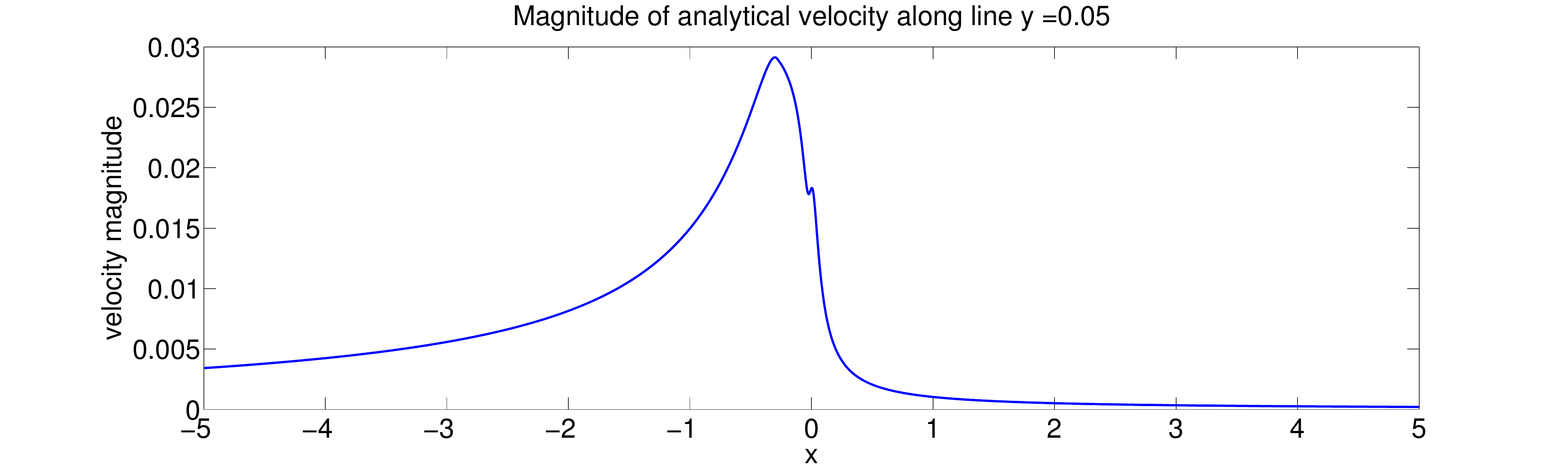}
    \caption{\figtext{Velocity boundary function at $\delta = 0.05$ for the third test problem where $\phi=170$, $U =0.02$, $Q=1$.}}
    \label{fig:BC_FN32}
\end{figure}

In the third and final model problem, the interface shape is assumed to be in the shape of a circular arc throughout the simulation. The Navier--Stokes part is again included. A circular interface shape is realistic for example when a liquid is rising in a narrow tube (capillary rise). 
For the third model problem we perform two set of simulations. In the first set the wall contact angle is $\phi = 140$ and in the second the angle is $\phi = 170$. Again we hypothetically relate these angles to a contact point velocity of $U = 0.02$. Just as before, we choose $\delta = 0.05$. The resulting velocity boundary function for the third test problem with contact angle $\phi = 140$ is the same as for the second model problem. The velocity boundary function for $\phi = 170$ is given in \figref{fig:BC_FN32}.

\begin{figure}[h!]
  \centering
    \includegraphics[width=0.95\textwidth]{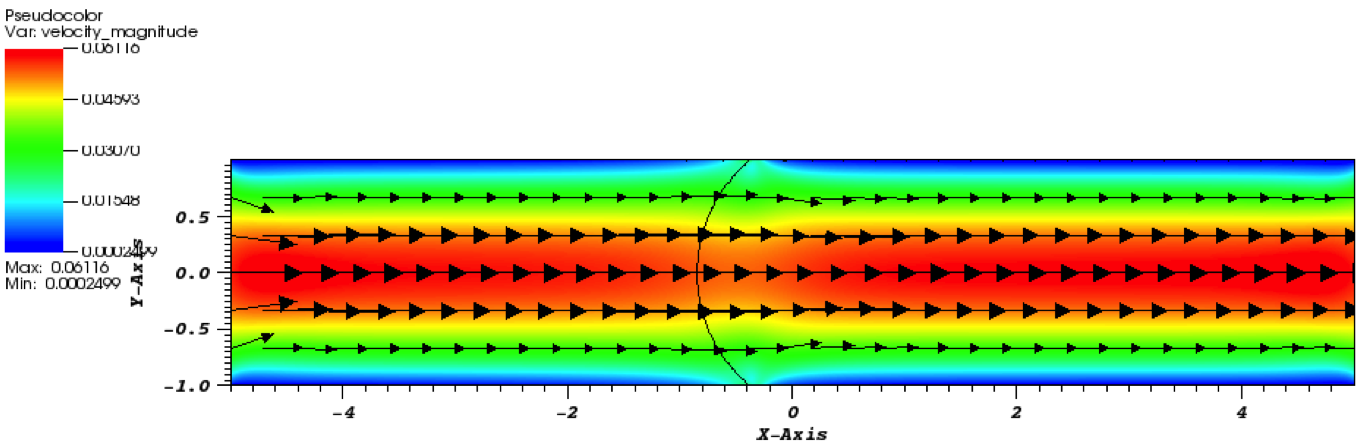}
    \includegraphics[width=0.95\textwidth]{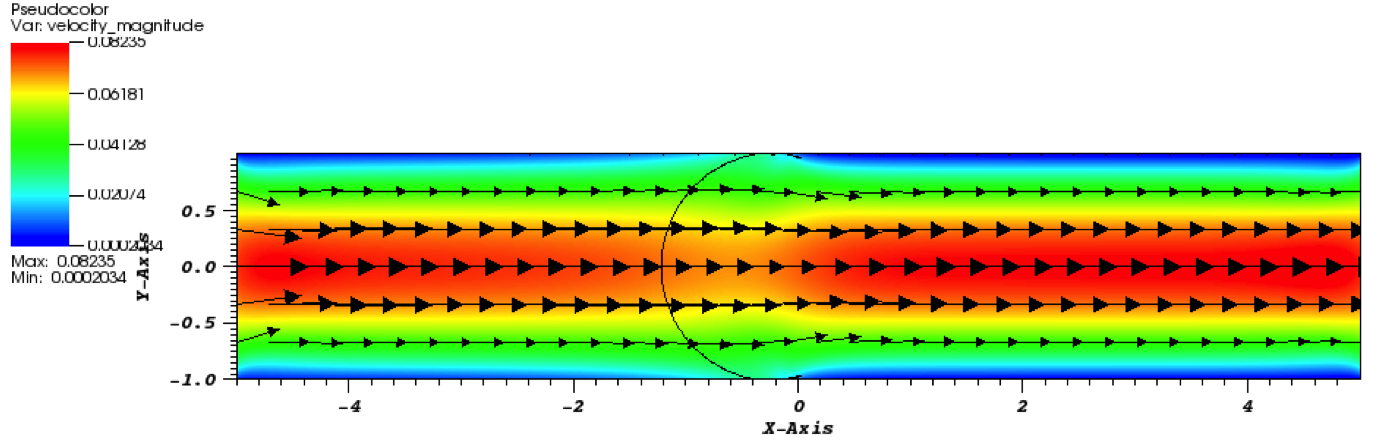}
    \caption{\figtext{Resulting velocity fields for the third test problem: $\phi = 140$ (upper) and $\phi = 170$ (lower), $U =0.02$, $Q=1$.}}
    \label{fig:result3}
\end{figure}

All grid sizes, except $h = 1/24$ for the case with a contact angle $\phi = 170$, are used and the resulting velocity fields at $T = 6$ for the grid size $h = 1/56$ is plotted in \figref{fig:result3}. It can be seen that away from the interface the flow profile is a regular poiseuille profile with zero velocity at the boundary. Close to the interface and the contact line however, the velocity is non-zero at the wall due to the velocity boundary condition. For this test problem (as opposed to the second test problem), it is clear that the fluid interface also effects the velocity field (via the surface tension).

\begin{figure}[h!]
  \centering
    \includegraphics[width=0.48\textwidth]{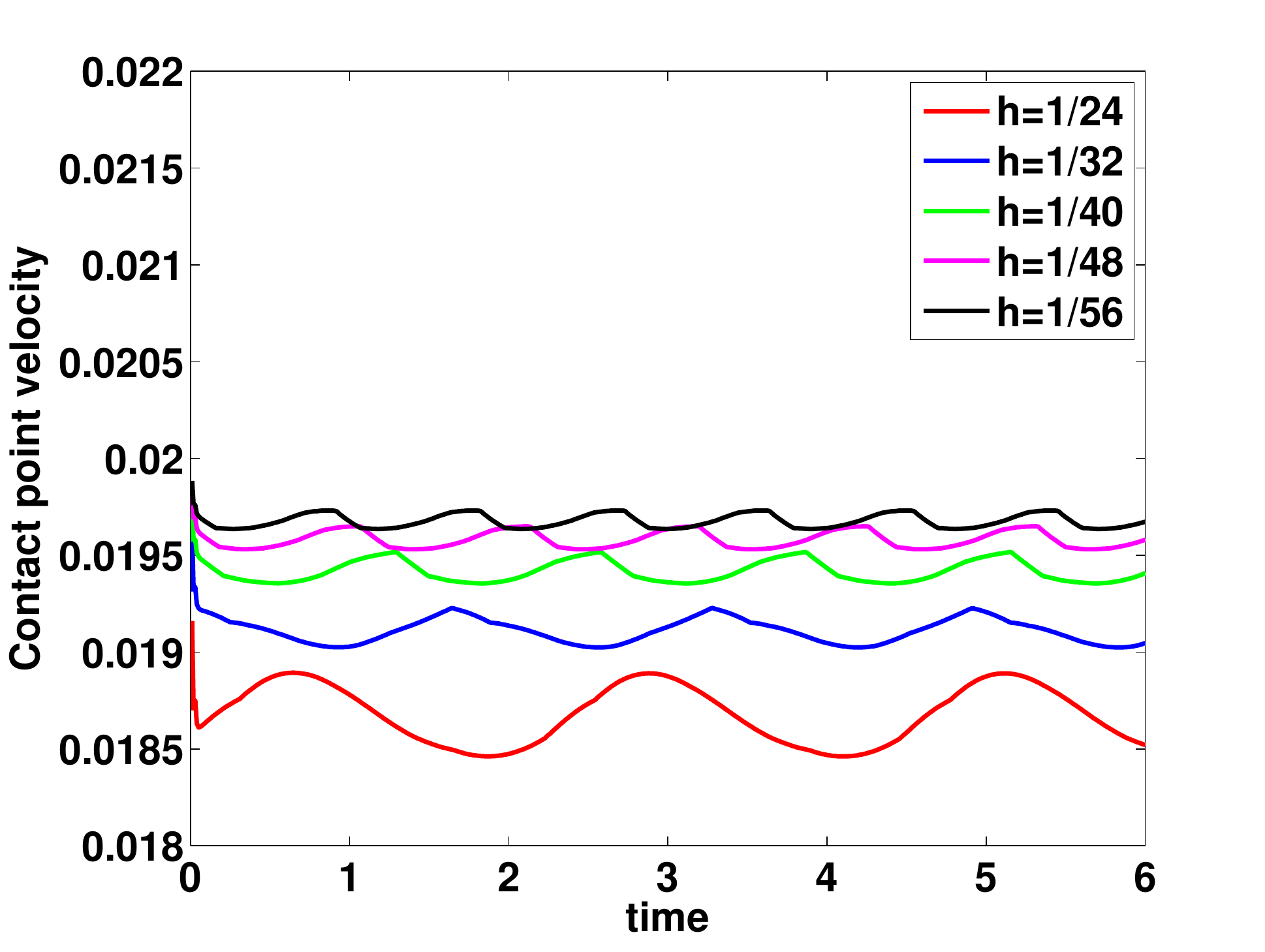}
    \includegraphics[width=0.48\textwidth]{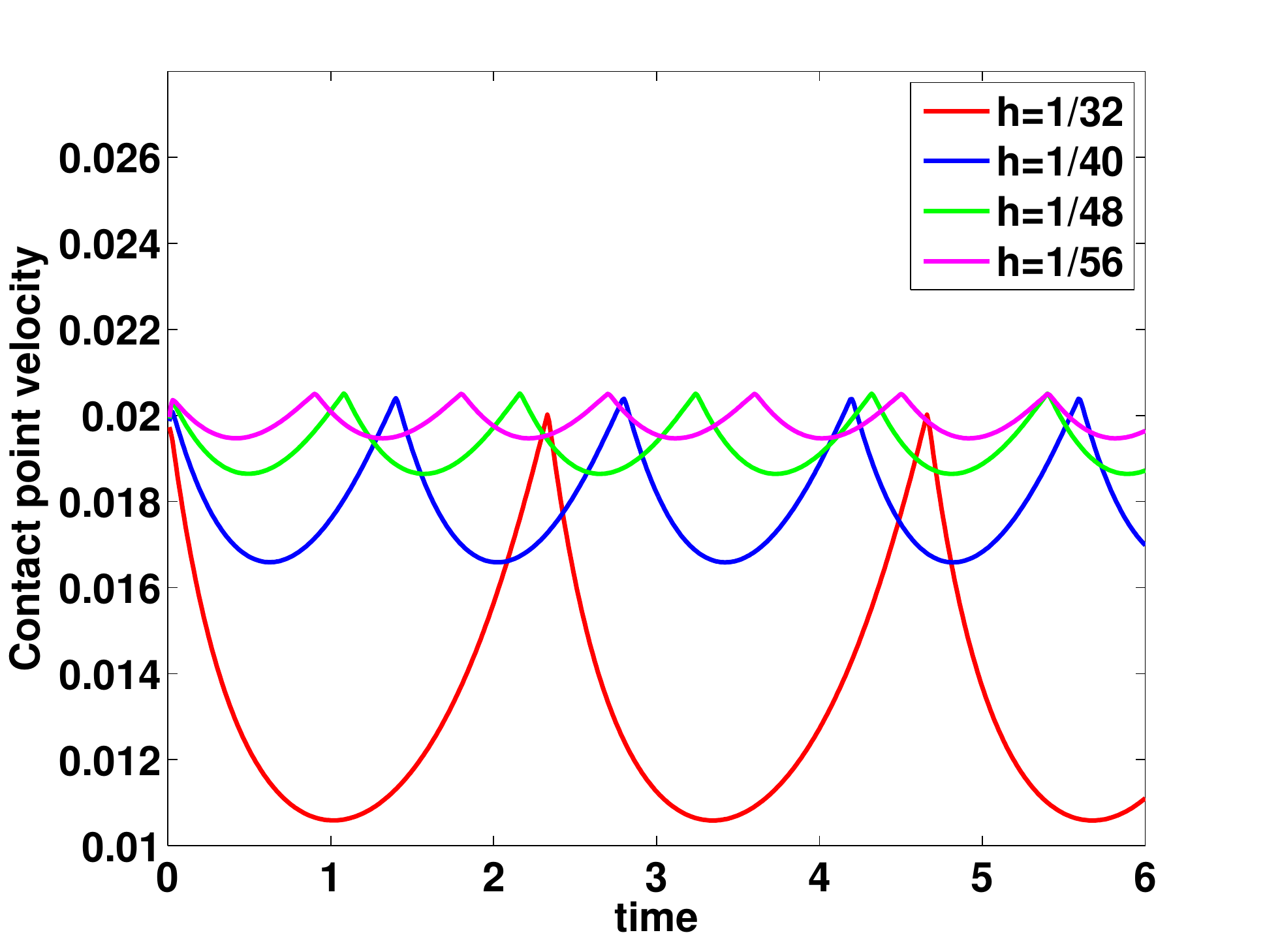}
    \caption{\figtext{Resulting contact point velocities for the third model problem, $\phi = 140$ (left) and $\phi = 170$ (right).}}
    \label{fig:velocities3}
\end{figure}

\begin{figure}[h!]
  \centering
    \includegraphics[width=0.48\textwidth]{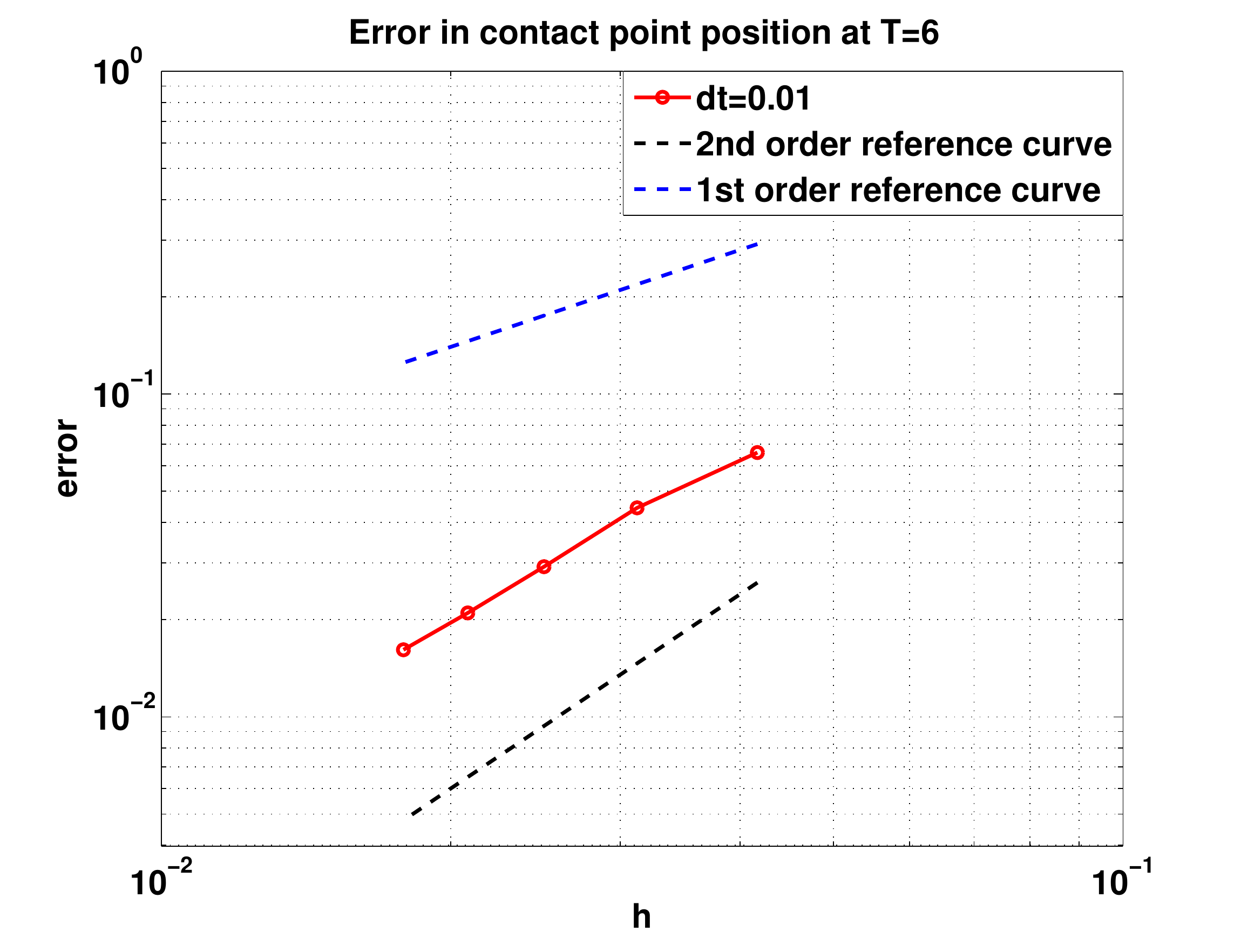}
    \includegraphics[width=0.48\textwidth]{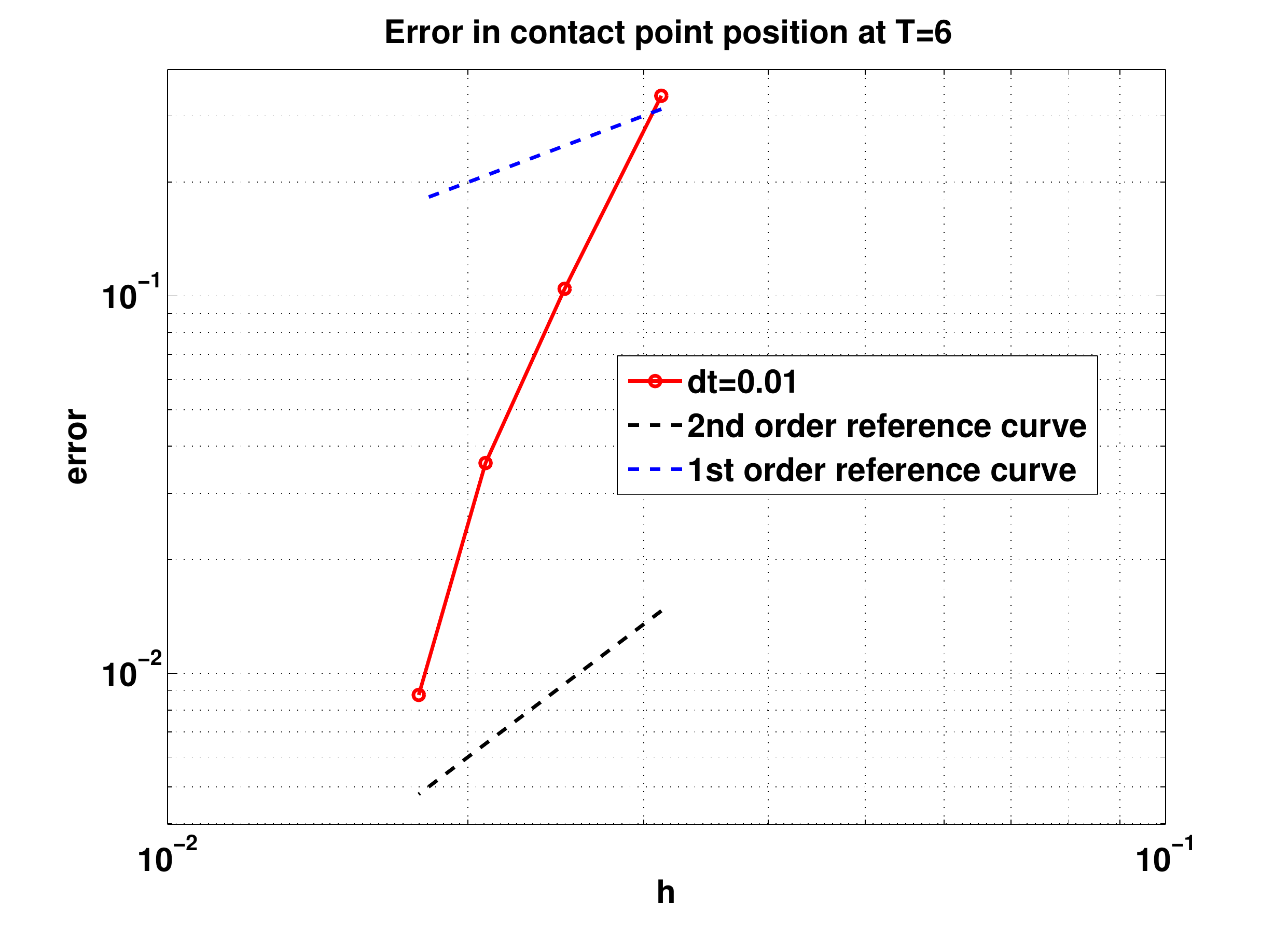}
    \caption{\figtext{Error in contact point position at T = 6 for the third model problem, $\phi = 140$ (left) and $\phi = 170$ (right). }}
    \label{fig:conv3}
\end{figure}

\newpage
The resulting contact point velocity in each time step is plotted as a function of time in \figref{fig:velocities3} and in \figref{fig:conv3}. It can be seen that grid convergence is again obtained with a rate of convergence $p \approx 2$ for the case where $\phi = 140$ and with rate of convergence $p > 2$ for the case where $\phi = 170$. The analytical model in \secref{sec:intermediate} is valid for contact angles $0 < \phi < 180$, but the case where $\phi = 170$ approaches the upper limit. This can be seen in that the errors are larger for that case than for $\phi = 140$, for all grid sizes except $h=1/56$. The high rate of convergence for the case with the larger contact angle is probably due to that the courser grids are not capable of resolving the problem well enough.  
\vspace{-0.3cm}

\section{Summary and Conclusions}
\label{sec:sum}
We present the first steps in constructing macroscopic velocity boundary conditions to impose a contact point velocity. The contact point velocity is assumed to be given as a result from microscopic models. The idea is based on introducing an intermediate scale where the creeping flow approximation is valid. This approximation is used to derive macroscopic velocity boundary conditions along a fictitious boundary, located a small distance inside the physical boundary. 

Model problems where the shape of the interface is constant thought the simulation are introduced. For these problems, experiments show that the errors in the resulting contact line velocities converge with the grid size $h$ at a rate of convergence $p\approx 2$. The result is sensitive to contact angles approaching the limit of $\phi=0$ or $\phi=180$, and for such cases a smaller grid size is needed to properly resolve the problem. Further, for an accurate implementation of the velocity boundary condition, it is important to accurately evaluate the contact line position in each time step. Here we use cubic interpolation.     

To be able to simulate problems with dynamic interface shapes, the model needs to be extended with a reinitialization of the level set function that preserves the contact line position (see \secref{sec:LS}). One possible approach is to implement the reinitialization developed in \cite{KUZMIN} where a penalty term is introduced to preserve the interface shape.

Further, to improve the model an approach where the computational boundary is modified only close to the contact line should be considered. Also, curved solid boundaries should be taken into account. 




\clearpage

\bibliography{references}
\bibliographystyle{ieeetr}

\end{document}